\documentclass[epj]{svjour}
\usepackage{latexsym}
\usepackage{amsmath}
\usepackage{amssymb}
\usepackage{graphicx}
\usepackage{subfigure}
\begin{document}
\title{Phase Synchronization in Railway Timetables}
\author{Christoph Fretter$^1$, Lachezar Krumov$^2$, Karsten Weihe$^2$, Matthias M\"uller-Hannemann$^1$, \and Marc-Thorsten H\"utt$^3$
\thanks{supported by Volkswagen Foundation grants I/82717, I/82718 and I/83435. The authors
  wish to thank Deutsche Bahn AG for providing test data for scientific use.}%
}                     % Do not remove

\authorrunning{Fretter, Krumov, Weihe, M\"uller-Hannemann and H\"utt}
\titlerunning{Phase Synchronization in Railway Timetables}

\institute{ Institut f\"ur Informatik, Martin-Luther-Universit\"at Halle-Wittenberg, Halle, Germany \and Fachgebiet Algorithmik, Technische Universit\"at Darmstadt, Darmstadt, Germany \and School of Engineering and Science, Jacobs University, Bremen, Germany}
\date{Received: date / Revised version: date}
\abstract{
%Train connection timetables are an important research topic in
%algorithmic. 
Timetable construction belongs to the most important optimization problems in
public transport.
Finding optimal or near-optimal timetables under the subsidiary
conditions of minimizing travel times and other criteria is a targeted contribution to the functioning of public transport. In
addition to efficiency (given, e.g., by minimal average travel times),
a significant feature of a timetable is its
robustness against delay propagation. Here we study the balance of
efficiency and robustness in long-distance railway timetables (in particular the current long-distance railway timetable in Germany) 
from the perspective of synchronization, exploiting the fact that a major part of the trains run nearly periodically. We find that synchronization is highest at intermediate-sized stations. We argue that this synchronization perspective opens a new avenue towards an understanding of railway timetables by representing them as spatio-temporal phase patterns. Robustness and efficiency can then be viewed as properties of this phase pattern. 
\PACS{
{89.75.Fb} 	{Structures and organization in complex systems} \and
{89.75.Hc} 	{Networks and genealogical trees} \and
{89.20.Ff} 	{Computer science and technology} \and
{89.40.-a} 	{Transportation} 
%      {PACS-key}{discribing text of that key}   \and
%      {PACS-key}{discribing text of that key}
     } % end of PACS codes
} %end of abstract
\maketitle

\section{Introduction}
Railway timetables should be designed to achieve a maximum level of
utilization from a passenger's perspective. 
That is, regular waiting times for connecting trains should
be kept to a minimum. However, this limits the network\rq{}s robustness
against perturbations: Depending on the waiting policy among
connecting trains,
a single delayed train may cause a cascade of
further train delays in remote parts of the network.
Minimal regular waiting times (minimal buffering times)
cause maximal risk of such delay propagation. Understanding this trade-off and
limiting the propagation of delays through the networks is a
challenge of practical importance.

\begin{figure}[b]
\centerline{\includegraphics[width=0.45\textwidth]{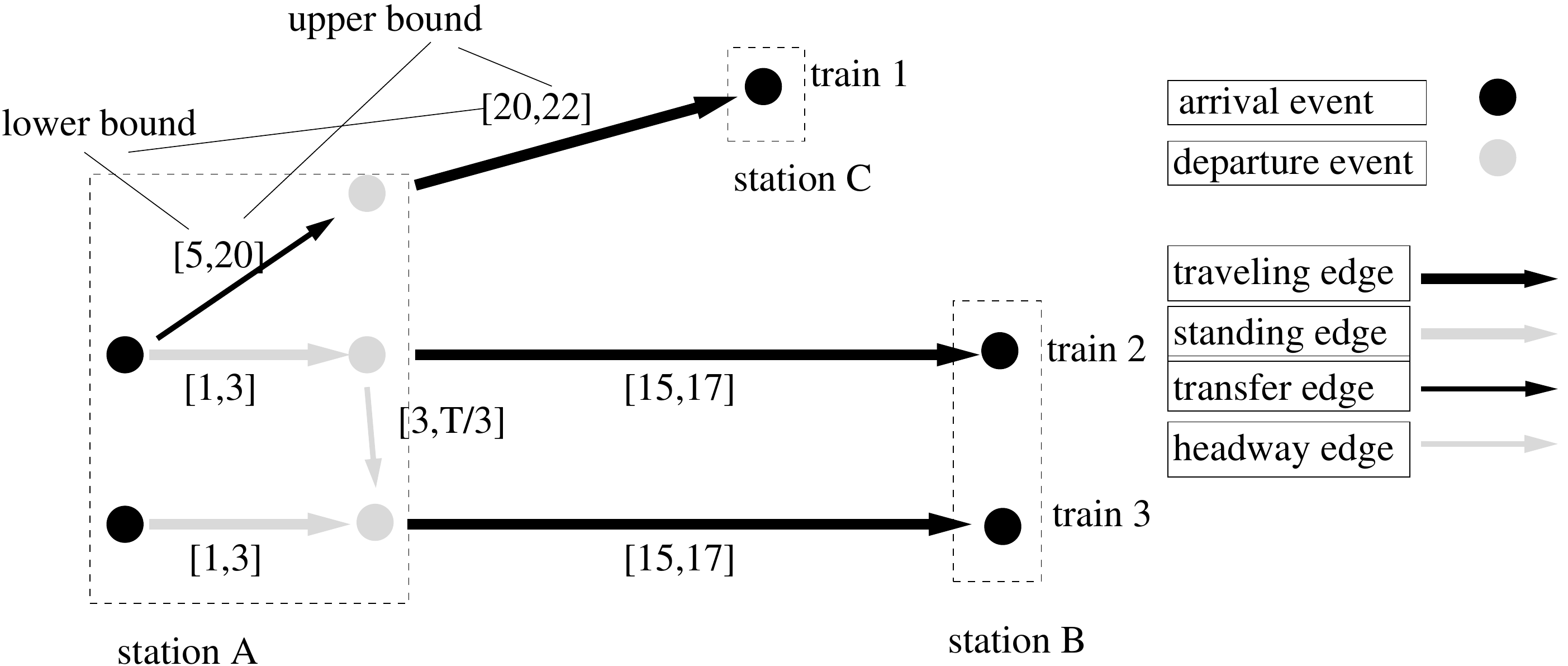}}
\caption{\label{fig:schedule-example}Small excerpt of a periodic event
  scheduling problem.}
\end{figure}

The construction of periodic railway timetables is algorithmically
difficult and has been intensively studied
as a periodic event scheduling problem (PESP), see for 
example~\cite{SerafiniUkovich89,Peeters03,Liebchen06}.
The technical and economical side constraints for a valid non-periodic
schedule
can be modeled as a feasible differential problem on a directed 
graph $G=(V,E)$ with lower and upper edge bounds $\ell,u \in \mathbb{Q}^E$. 
In a basic model,
the vertex set $V$ corresponds to departure and
arrival events, while the directed
edges together with the bound values 
model constraints (travel times, minimum headway, minimum
transfer times, etc.).
One seeks for a vector
$\pi \in \mathbb{Q}^V$, called the timetable, 
which assigns to each event $j$ a time-stamp $\pi_j$
satisfying
$$ \ell_e \leq \pi_j - \pi_i \leq u_e \text{ for all } e =(i,j) \in E.
$$ 
Thus, lower and upper edge bounds restrict the difference 
between two timestamps from below and above, respectively.
For example, $\ell_{(i,j)} = 15 \leq \pi_j - \pi_i$ means  that
event $j$ has to occur at least 15 time units after event $i$.
See Figure~\ref{fig:schedule-example} for a small example.

In a periodic timetable, trains are grouped into lines which are to
be operated by some period $T$.
In the periodic event scheduling problem
(with one fixed period $T$)
one searches for a
vector $\pi \in [0,T)$ such that
for all $e =(i,j) \in E$ there exists $k_e \in \mathbb{Z}$ with
$$\ell_e \leq \pi_j - \pi_i + T \cdot k_e \leq u_e .$$ 
For the local public transport in Berlin (Germany), the first optimized periodic
timetable used in daily operation has been obtained using
mixed-integer linear programming techniques~\cite{Liebchen08}. 
Netherlands Railways also have recently introduced a completely new 
periodic timetable, generated by a
number of sophisticated operations research techniques, including
constraint programming~\cite{Kroon-et-al2009}.
%This approach does not generalize to larger systems (e.g., long-distance train connections) or to systems with less strict periodicity. 
For countries with a less periodic timetable, including Germany,
the construction process for long-distance timetables is quite
complex, and therefore still done to a large extent manually 
by experienced engineers. 
The planning process has a hierarchical component (international trains are
scheduled first), and a behavioral component (keep as much as possible
from the previous year's schedule).

So far, railway timetables have been studied predominantly as an algorithmic 
challenge with the objective of constructing optimal (or near-optimal) connection patterns, minimizing resources and overall waiting time. 
Only recently, there have been first computational studies aiming at
delay resistant periodic 
timetables~\cite{Kroon-et-al07,Liebchen-et-al10,LiebchenStiller09}.

Here we adopt an opposite perspective to timetable construction and analyze the spatio-temporal patterns induced by the timetable. 
A suitable language for this study is a representation of the train arrival/departure events as a spatio-temporal phase pattern. We study the distribution of synchronization across stations.
Synchronization phenomena have received a lot of attention in traffic
modeling over the last few years, in particular for car traffic in
cities and the impact of traffic light synchronization on the
formation of traffic jams
\cite{PhysRevE.64.056132,1742-5468-2008-04-P04019,Lämmer200639}.

 In the case of railway timetables, the situation is different in
 several ways: The ``load'' of a station is essentially determined by
 the number of tracks (giving the maximal number of simultaneous or
 nearly simultaneous arrival/departure events). The typical number of directions
 (which can be interpreted as the degree of a station in a suitable 
 effective network representation), from which one can select, is
 higher for train stations than for typical street crossings.

% The step from the railway timetable to a dynamical process on a network
% requires deciding on a suitable network representation. In our
% analysis, nodes are pairs consisting of a train $tr_j$ and station
% $S_k$, while links are travel events from one station to another or
% train-change events at a station. 
% Many other network representations of the railway timetable are possible. In our representation trains at different times are collapsed onto this time-integrated network. One could for example include time as an additional parameter characterizing a node and thus look at effective networks within certain time intervals. Relatedly, one could consider stations as nodes and draw a link between two stations, when the one station can be reached from the other within a certain time interval (reachability network). Another set of alternative network representations uses the routes through the system (i.e. the actual traveled trajectories) as a starting point and construct a route-based network from alternative paths. 

If one considers a network of long-distance train connections as a
mesh of routes through a planar system, where trains are started
periodically at the endpoints of these routes, the spatial distances
between the intersection sites of these routes determine a
spatio-temporal phase pattern. The free parameters of this pattern are
the relative phases of the periodically started trains. In reality,
the travel time between two stations can serve as an additional degree of freedom 
allowing for a shaping of the phase pattern beyond this simple thought
experiment. 

Our main hypothesis is that the rank of the stations sorted according
to size is the organizing parameter (i.e. the ``control parameter"
from the perspective of self-orga\-nized systems \cite{mikhailov,kurths,walleczek}), along which synchronization can be understood.
%
%We therefore perform averages over neighboring ranks in order to reduce fluctuations in the data. 
%
In this paper, we use the notion 
buffering time to denote  
the amount of time available to change between two trains
(transfer time) for the planned schedule, i.e. without induced delays. 
Our other two observables are the average buffering time $b_i$ at
station $i$ and the secondary delays $s_i(p)$ induced by a primary
(incoming) delay $p$ because trains have to wait for other trains.

The main result of our analysis is that a railway time\-table
% constitutes / 
induces %/ encodes 
a spatio-temporal phase pattern, and that properties of the phase pattern are linked to the efficiency and the robustness of the  system. 
 We observe that synchronization is highest at intermediate-sized stations.

Here we contribute two points to the general debate: 

(1) We show that the current planning of railway time\-tables (which involves some algorithmic construction, some manual curation and the resorting to features from previous timetables) leads to an unexpected coherence on the level of the spatio-temporal phase pattern. 

(2) At the same time, our analysis shows that the general concept of a spatio-temporal phase pattern is a novel and helpful view for network-based scheduling problems.

The remainder of this paper is structured as follows. 
In Section~\ref{sec:formalism}, we first give a
detailed description of our numerical experiment,
and then discuss the results in Section~\ref{sec:results}. 
Afterwards, we introduce an avalanche model for delay propagation on graphs (Section~\ref{sec:avalanche}) helping us to understand the observed relation between synchronization and robustness.
Finally, we conclude with a short summary and an outlook for future work (Section~\ref{sec:summary}).

\section{Formalism and Numerical Experiments}\label{sec:formalism}

The quality of the time\-table
is related to two distinct (and often conflicting) objectives: 
%/ requirements: 
The sum of travel times over all routes should be minimal (efficiency)
and typical delays should minimally increase the overall travel
time (robustness). 
Apart from some freedom to determine the
planned travel time from one station to another (i.e. the prescribed average speed of the train), the main tuning capacity lies in the interchange time between connecting trains. While efficiency requires a minimization of interchange time, robustness can  be established by using the interchange time as a buffer for incoming delays.

The secondary delays $s_i$ observed at
each station $i$ across a range of primary delays $p$ 
have been obtained by a large-scale numerical
experiment performed on the actual timetable of Deutsche Bahn AG, 
together with real passenger information.
Throughout our investigation we consider only long-distance train
connections (served by the train categories ICE and IC/EC). 
To simulate the effects of delays, we use the dependency graph model 
introduced in~\cite{Mueller-HannemannSchnee09} and its implementation
within  the fully realistic multi-criteria timetable information 
system \mbox{MOTIS}~\cite{Schnee09}. 
The dependency graph is basically a time-expanded graph model with
distinct nodes for each departure and arrival event in the entire schedule
for the current and following days. In addition, the model includes
two further types of nodes: forecast and schedule nodes.

\begin{figure}[t]

\vspace*{-1cm}

\centerline{\includegraphics[width=0.45\textwidth]{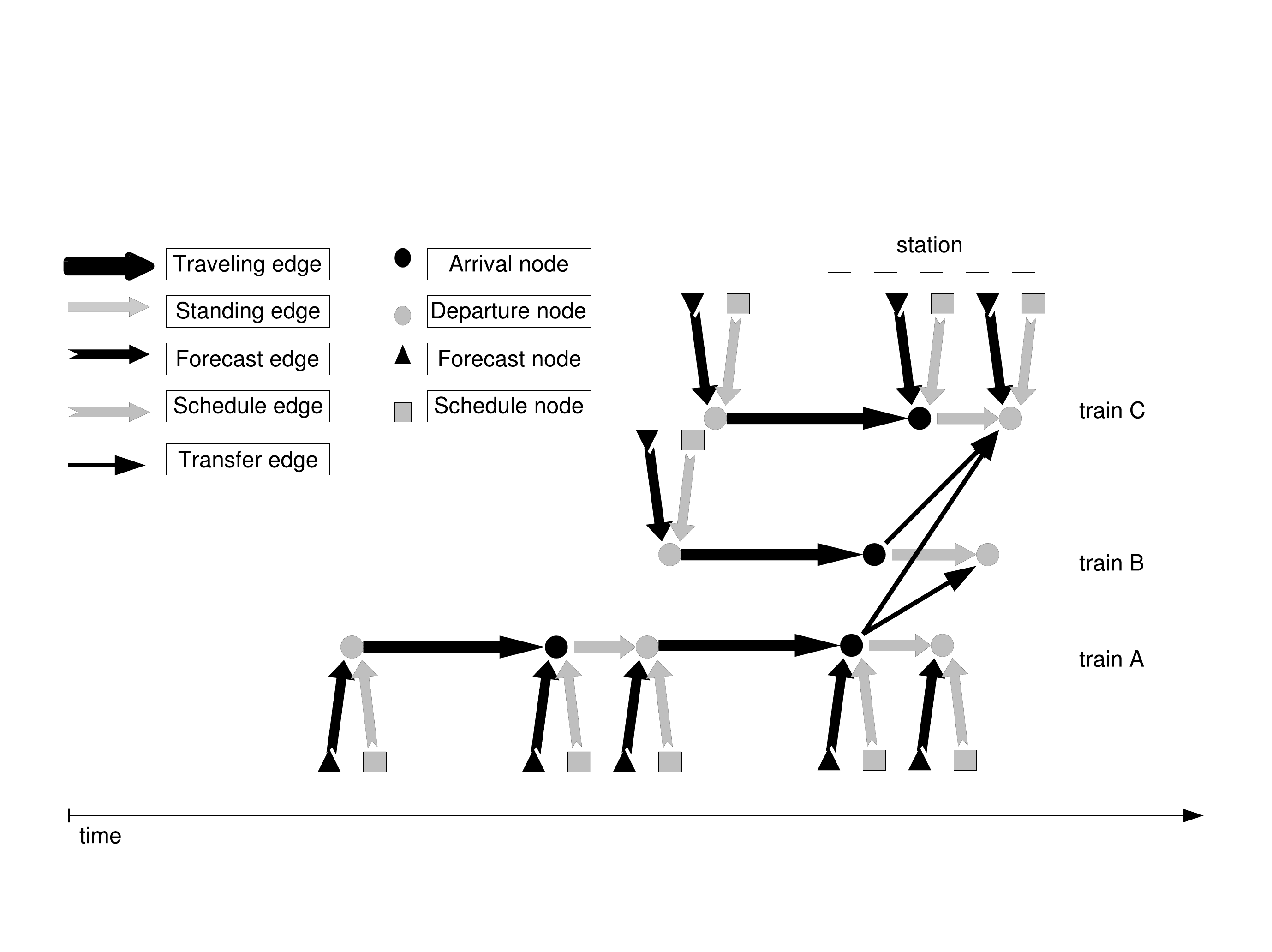}}

\caption{\label{fig:dependencygraph} Illustration of the dependency 
graph model (taken from~\cite{Mueller-HannemannSchnee09}).
}
\end{figure}

Each node has a time-stamp which can dynamically change. 
The timestamps reflect the current 
situation, i.e.\ the expected departure or arrival time 
subject to all delay information known up to this point.
Schedule nodes are marked with the planned time of 
an arrival or departure event, whereas 
the time-stamp of a forecast node is the 
current external prediction for its departure or arrival time.

The nodes are connected by five different types of edges 
(see Figure~\ref{fig:dependencygraph}). The purpose of an
edge is to model a constraint on the time-stamp of 
its head node. 
\begin{itemize}
\item
\emph{Schedule edges} connect schedule nodes to departure 
or arrival nodes. They carry the planned time for the corresponding event 
of the head node (according to the published schedule).  
\item
\emph{Forecast edges} connect forecast nodes to departure 
or arrival nodes.
They represent the time stored in the associated forecast node.
\item
\emph{Standing edges} connect arrival events at a certain station 
to the following departure event of the same train.
They model the condition that the arrival time of 
train $t$ at station $k$ plus its minimum standing time 
must be respected before the train can depart (to allow for boarding 
and disembarking of passengers).
\item
\emph{Traveling edges} connect a departure node of some train $t$ 
at a certain station $k$ to the very next arrival node of this train 
at station $k'$. 
\item
\emph{Transfer edges} connect arrival nodes to departure nodes 
of other trains at the same station, if there is a planned transfer 
between these trains.
\end{itemize}

The current time-stamp for each departure or arrival node can be
defined recursively, for details see~\cite{Mueller-HannemannSchnee09}.

The MOTIS tool can be used as follows. Given the planned train
connection of a passenger and a concrete delay scenario (for example, a
single primary delay of a train), we can query MOTIS for the fastest
train connection towards the passenger's destination, subject to the
standard waiting rules between connecting trains. 
In particular, the train waiting regulations of Deutsche Bahn have been used. 
From the difference
between the planned arrival time at the destination and the calculated
arrival time in the delay scenario we obtain the individual delay for
each passenger.

\begin{figure}[t]
   \begin{center}
        \includegraphics[width=0.45\textwidth]{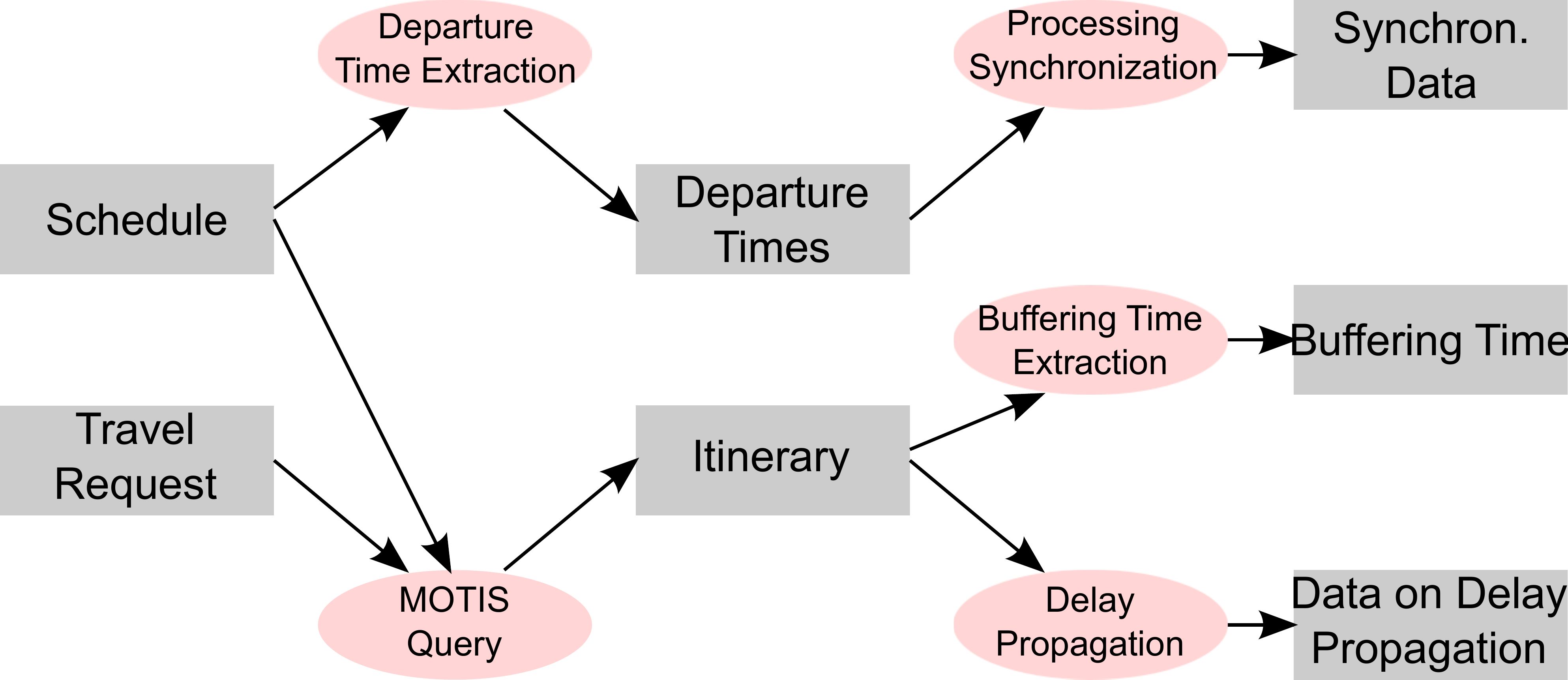}
        \caption{Data flow in the delay propagation experiment.\label{fig:flowChart}}
    \end{center}
\end{figure}

Passenger information has been available to us for a single day in the form of all travel agency bookings for that day. While these data are certainly distorted by the fact that most tickets are sold via vendor machines at the station (and these data have not been available to us), it is nevertheless helpful to include passenger data for two reasons: 

(1) Only routes, which have really been traveled, enter our analysis; in this way we avoid artifacts, e.g., from back-and-forth contributions.

(2) We can discuss both the average delay per passenger and the cumulative delay over all affected passengers (as a measure of the total systemic effect).

%It should be noted that the system is highly capsuled and, even on the level of numerical simulations, cannot be altered. It was therefore not possible to redo the same experiment at different waiting policies. 

In Figure~\ref{fig:flowChart}, we sketch
the data flow within our numerical experiment, where we have processed
$43 772$ train segments, $2622$ stations, $130 071$ passenger routes,
and about $1.8$ million
%$1 794 652$ 
MOTIS queries.

\begin{figure}[bt]
   \begin{center}
        \includegraphics[width=0.45\textwidth]{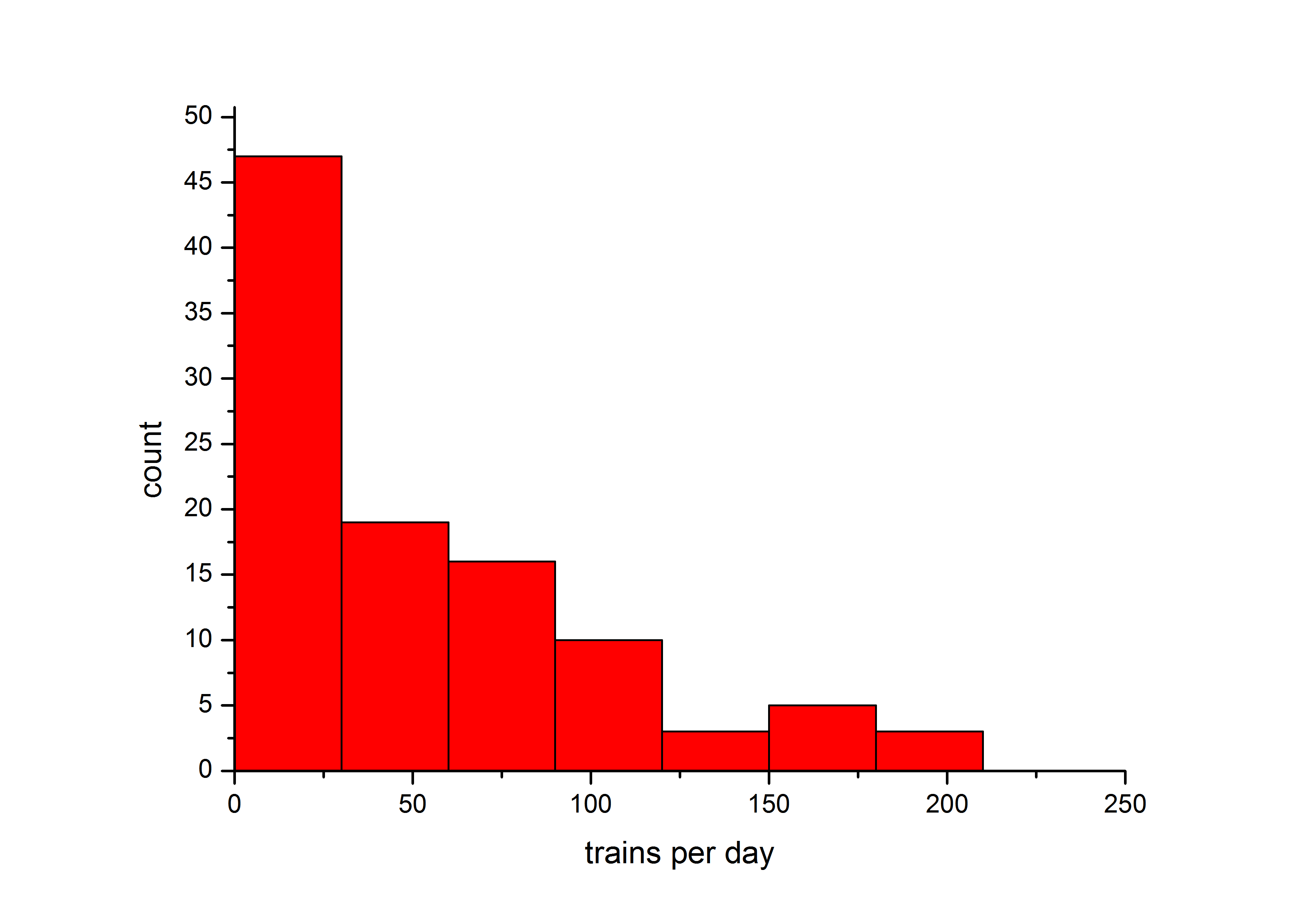}
        \caption{\label{fig:station_size}Distribution of daily numbers
          of arrival/departure (A/D) events.}
      \end{center}
\end{figure}

\begin{figure}[ht]
   \begin{center}
        \includegraphics[width=0.45\textwidth]{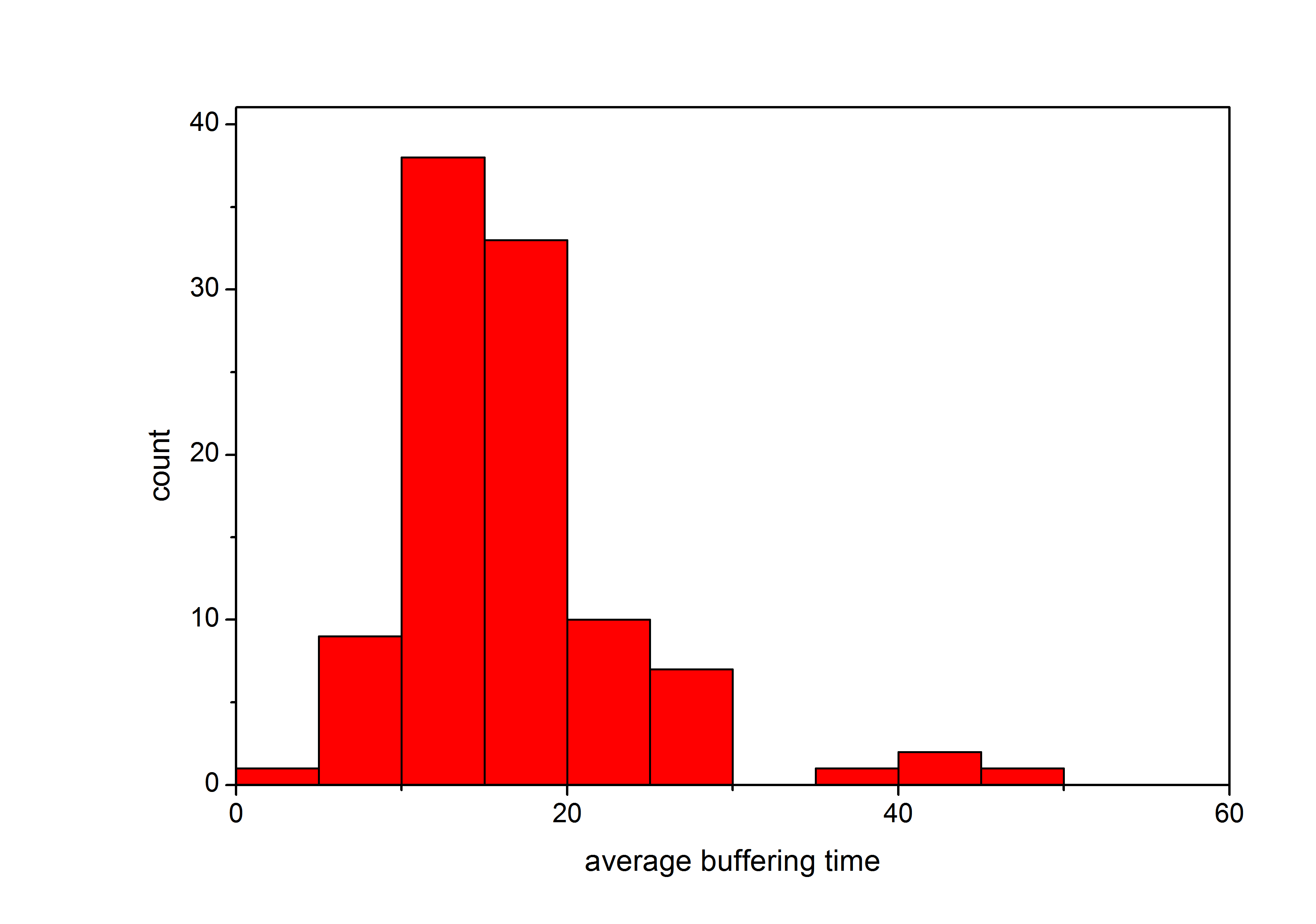}
        \caption{\label{fig:buffer_time}Distribution of the average buffering time per station.}
        
    \end{center}
\end{figure}

In order to illustrate the raw data obtained from this numerical
experiment, we show the station size distribution (where the station
size is given by the number of arrival/departure events per day) 
in Figure~\ref{fig:station_size}; and the buffering time distribution in Figure~\ref{fig:buffer_time}.
Both distributions are essentially unimodal and have a non-negligible
tail at large values. The rare occurrence of low buffering times can be
explained by the fact that the timetable information system does not
provide connections where a (station-specific) minimal interchange
time is not reached. 
It should be noted that this general rule is accompanied by a long list of exceptions for specific trains and specific connections. All these constraints and subsidiary conditions have been included in the numerical experiment, in order to obtain realistic event data. 

\begin{figure}[t]
\centerline{\includegraphics[width=0.5\textwidth]{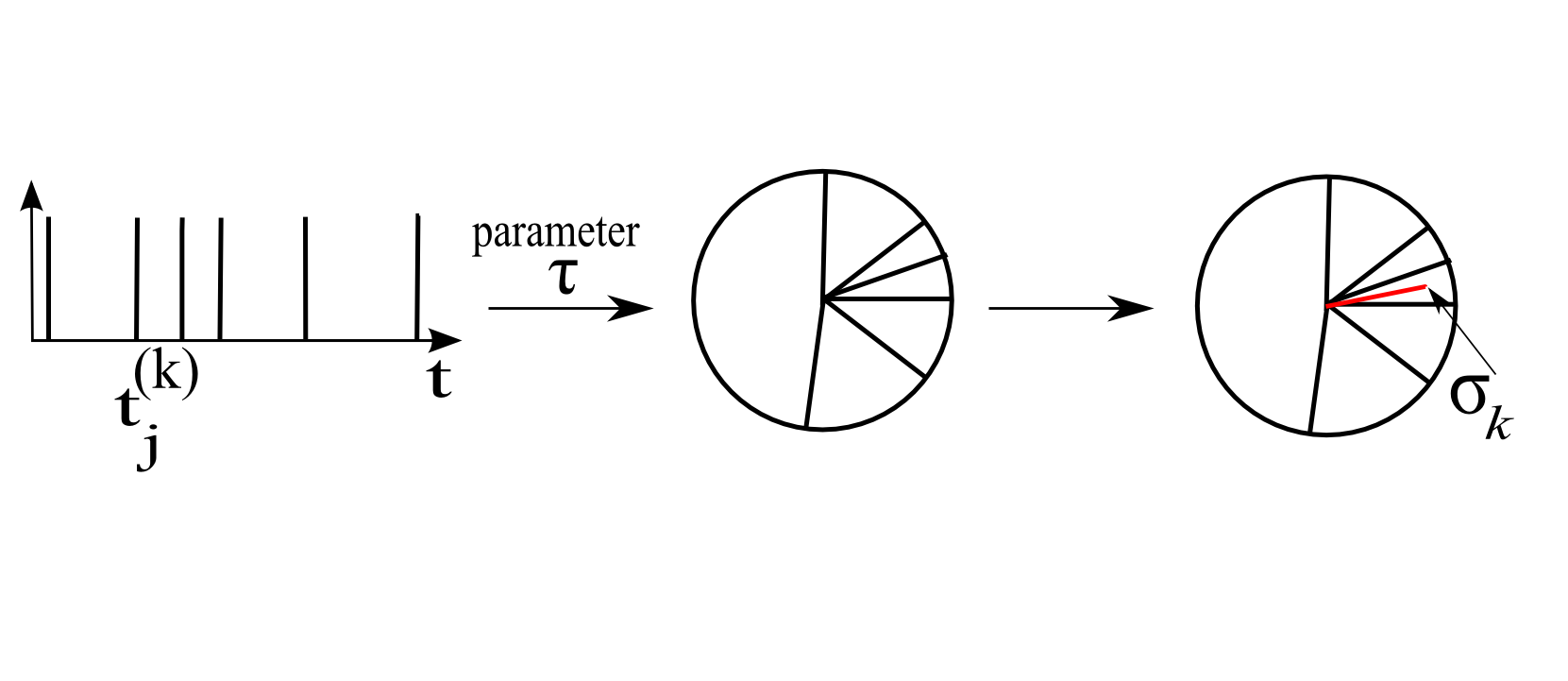}}
\caption{Conversion of the arrival/departure times at a station $k$
  into the synchronization index $\sigma_k$. \label{fig:sketch-of-synchronization}}
\end{figure}

As a next step we compare these delays with unperturbed features of the timetable. Our approach for converting the time\-table into an event pattern uses the language of phase synchronization.
Let  $\left\{ t_{j}^{(k)}, j = 1 \dots T_k \right\}$ be the set of arrival/departure (A/D) times $t_{j}^{(k)}$ of the $j$th train at station $k$. The quantity $T_k$ denotes the number of A/D events at station $k$ per day. These A/D times are now translated into phases

\begin{equation}
\phi_j^{(k)}(\tau) = \frac{2\pi}{r}(t_{j}^{(k)} mod \quad  \tau)
\end{equation}
with the period length $\tau $ as a parameter. In our analysis we set
this parameter to the maximal period length observed in the system,
i.e., $\tau $ = 120 minutes. For each station $k$ we can now compute
the synchronization index (as known from the classical studies of synchronization in populations of phase oscillators, see \cite{kuramoto,Strogatz2001di,winfree}; see also the scheme depicted in Figure~\ref{fig:sketch-of-synchronization}):

\begin{equation}
\sigma_k = \sigma_k(\tau) = \left|  \frac{1}{T_k} \sum_{j=1}^{T_k} e^{i \phi_j^{(k)}(\tau)} \right|
\end{equation}

The view we want to propagate here, is that the performance (in a very general sense) of a given timetable of train connections is related to its phase pattern.

\section{Results}\label{sec:results}
% secondary delay 

The large-scale numerical experiment described in the previous section in particular yields realistic
values of the secondary (induced) delay $s(p)$ as a function of the
primary (input) delay $p$. While at low $p$ the value of $s(p)$ is
mainly (but indirectly!) shaped by the buffering time $b$, at higher
$p$ the value is strongly influenced by the number of alternative connections.

\begin{figure}[t]
   \begin{center}
        \includegraphics[width=0.5\textwidth]{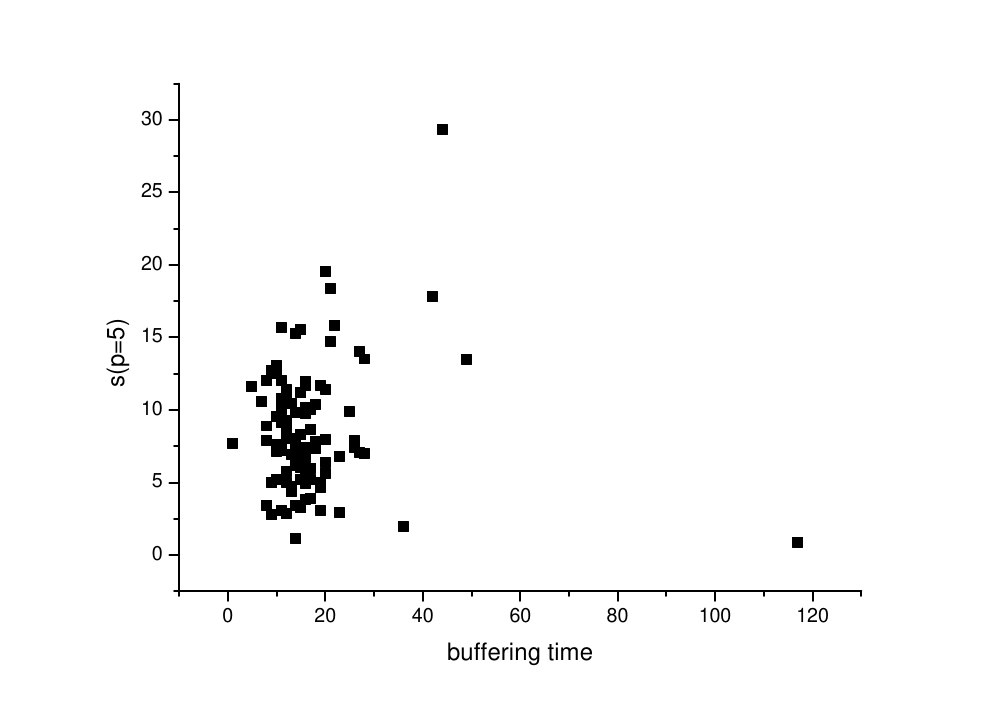}
        \caption{Secondary delay as a function of the buffering time for
          a fixed primary delay $p$ = $5$ minutes, raw data. \label{fig:secondary_delay_buffer_raw}}
    \end{center}
\end{figure}

\begin{figure}
\centerline{\includegraphics[height=5.5cm]{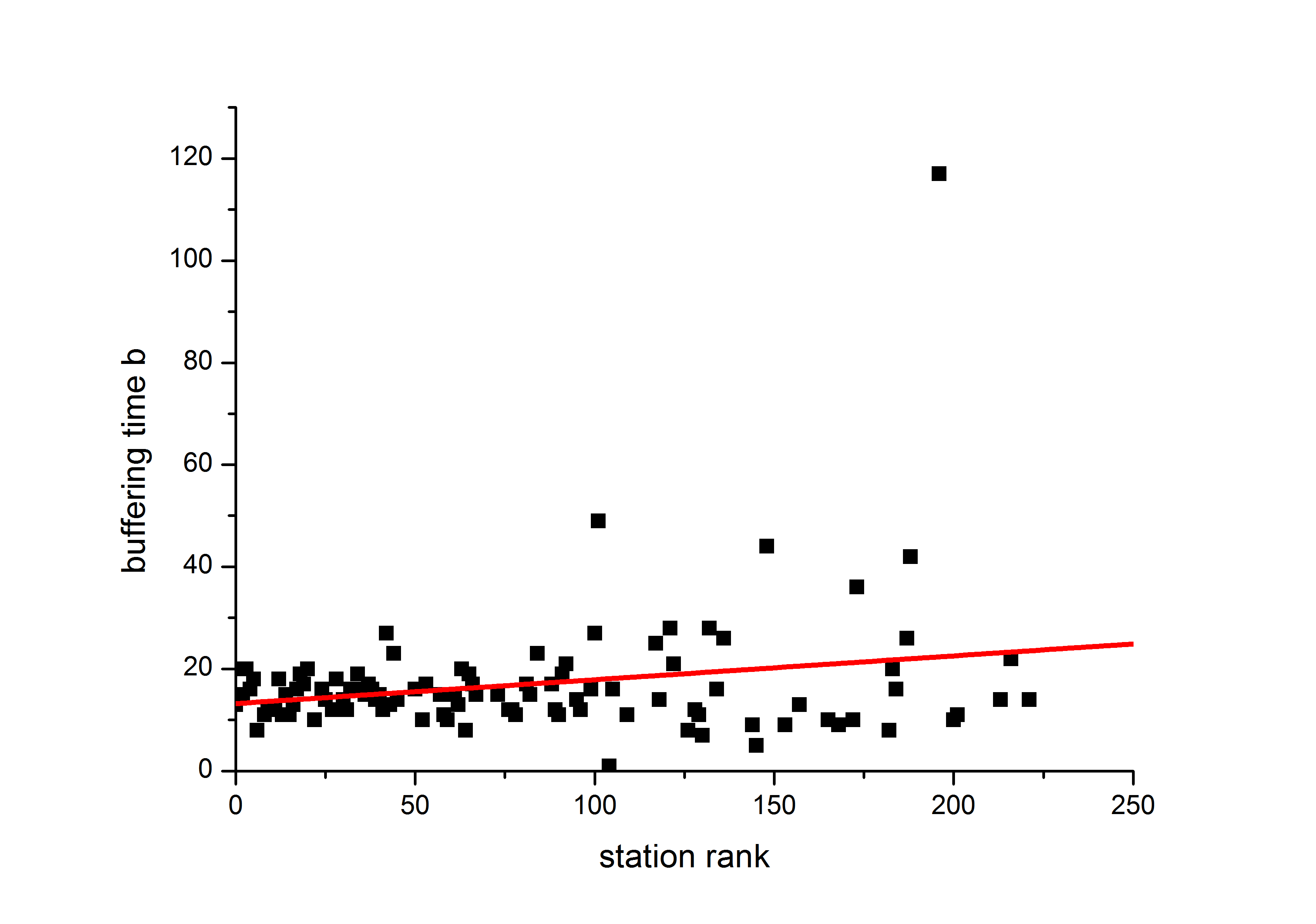}}
\caption{\label{fig:large_primary}Dependence of buffering time $b$ on
  station rank.}
\end{figure}

On face value, one would expect a negative correlation of the secondary delay $s(p)$ and the buffering time $b$ in this low-$p$ region.
In the raw data, Figure~\ref{fig:secondary_delay_buffer_raw}, there is
rather a lack of correlation (or even a slight tendency towards
positive correlations), which can be explained as follows: The
buffering time $b$ grows slowly with the station rank, i.e. decreases
slightly with the station size
(cf.\ Figure~\ref{fig:large_primary}). 
At the same time, larger stations (i.e. more A/D events) offer more alternative routes, effectively reducing the secondary delay, even at low primary delay $p$.

\begin{figure}
\begin{minipage}{0.5\columnwidth}
        \centerline{\includegraphics[width=\textwidth]{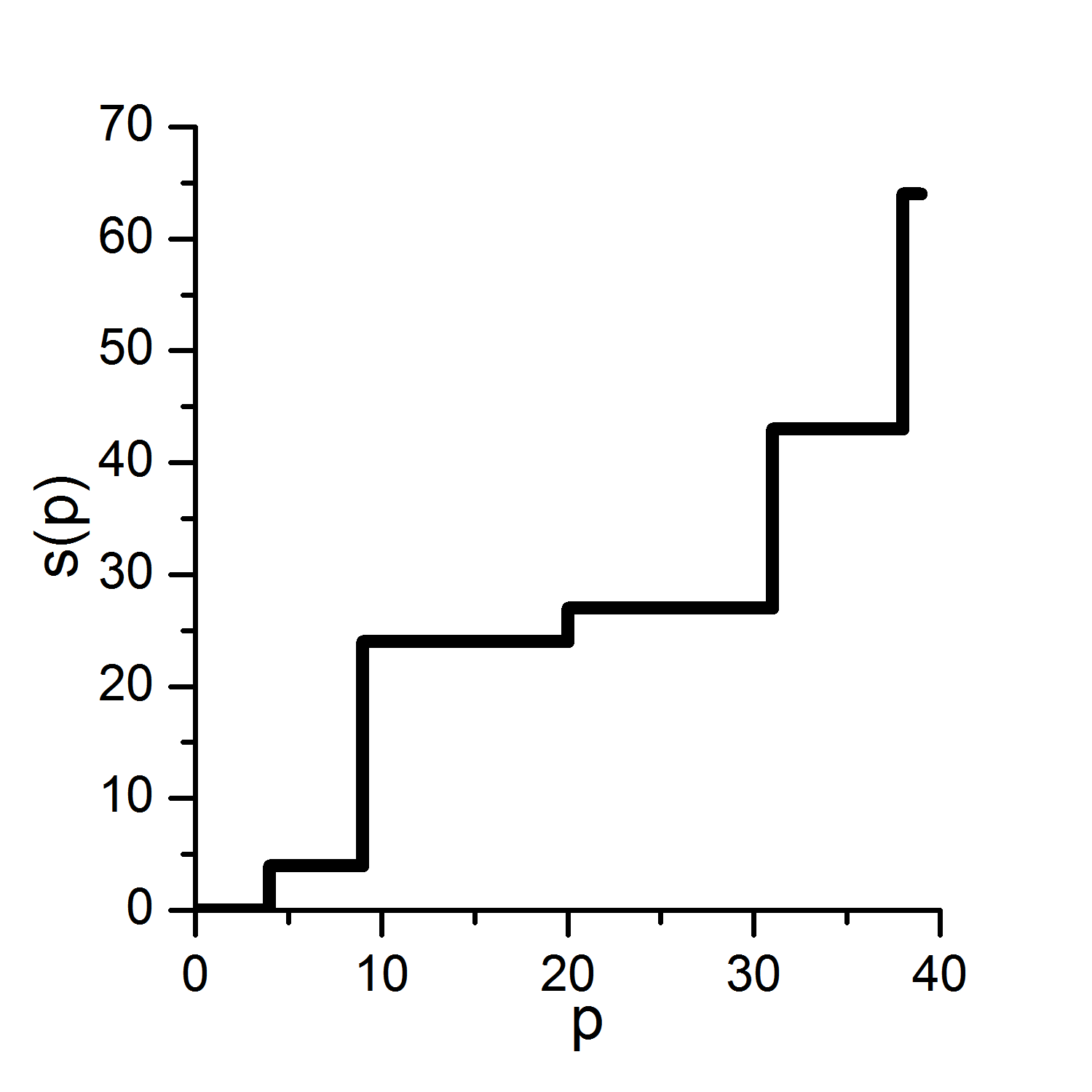}}

        \caption{Secondary delay as a function of the primary delay for a single train.  \label{fig:single_train}}

\end{minipage}
\hfill
\begin{minipage}{0.45\columnwidth}
\vspace*{3.5ex}

        \includegraphics[width=1.1\textwidth]{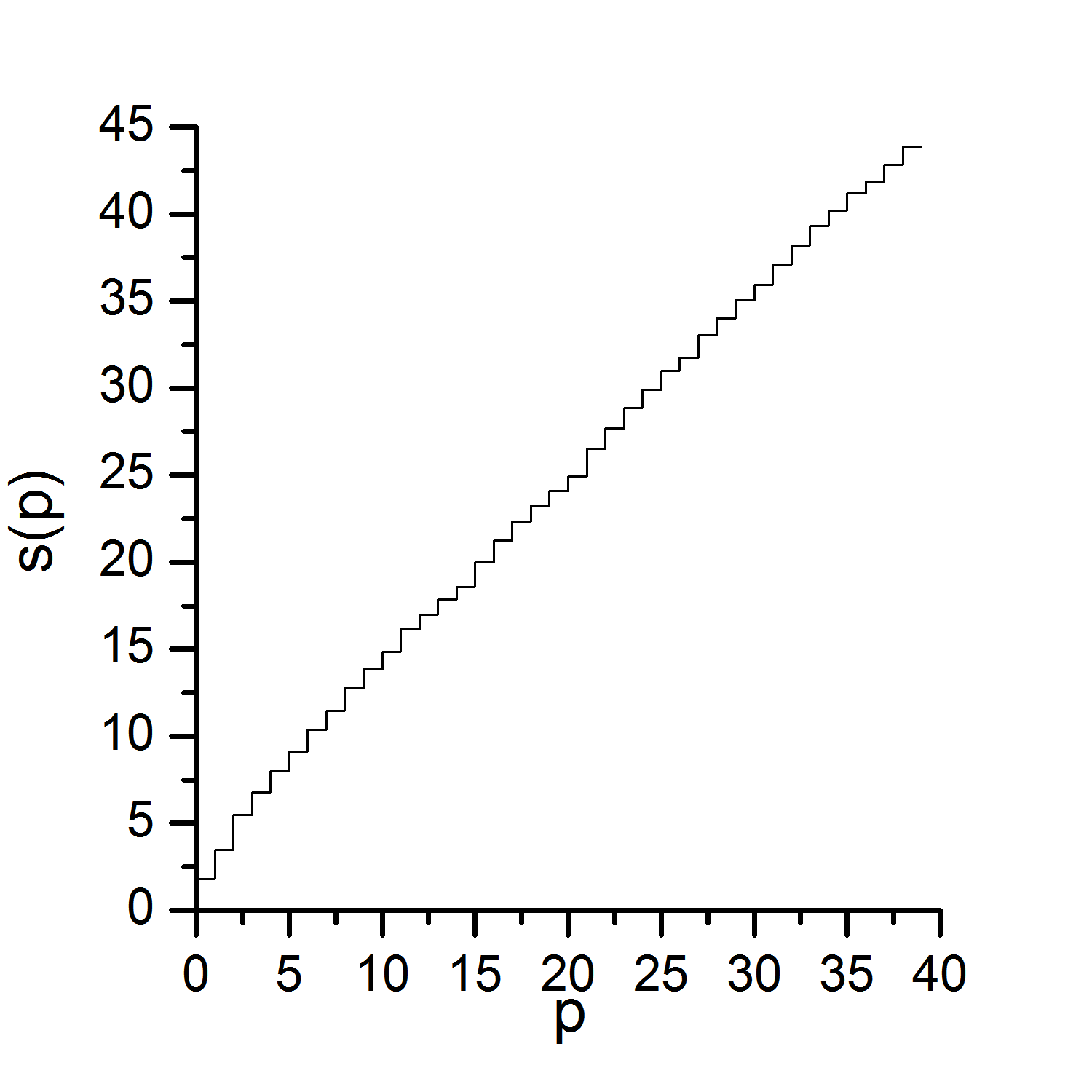}

\vspace*{2ex}

        \caption{Average secondary delay as a function of the primary
          delay for a single station (here: Frankfurt (Main) central).
          \label{fig:single_station}}
\end{minipage}
\end{figure}

In the example shown in Figure~\ref{fig:single_train},
there are only 4 minutes of buffering time which induce no delay.
When $p$ is in the range of $[4,9]$ minutes, the secondary delay
becomes $4$ minutes,
and then jumps to 25 minutes.
Figure~\ref{fig:single_station} shows the average secondary delay at a station, which is, as a first approximation a linear function.
At higher values of $p$, additional effects can be expected to set in:

(1)  with higher $p$ more alternative routes become 

accessible,

(2) more passengers will be affected, and 

(3) longer avalanches of delayed trains are triggered 

upon waiting. 

These contributions are partially compensated by the waiting policy: Avalanche length is strongly reduced by maximal waiting times. Also, the second contribution has a smaller (but still non-zero) effect on the average secondary delay per passenger.

\begin{figure}[ht]
   \begin{center}
        \includegraphics[width=0.5\textwidth]{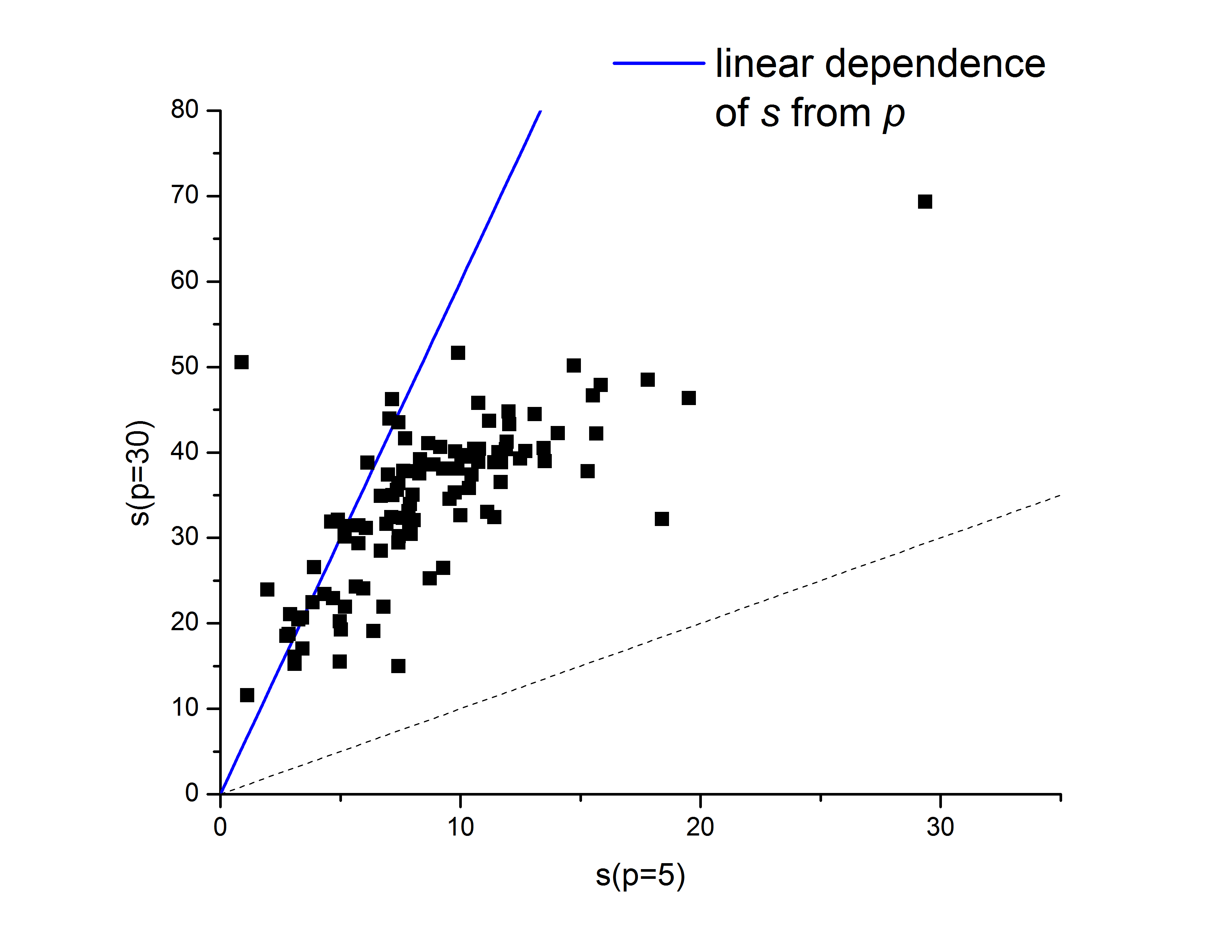}
        \caption{Correlation of the secondary delay with primary delay of 5 and 30 minutes.}
        \label{fig:primary_delay_correlation}
    \end{center}
\end{figure}

Figure~\ref{fig:primary_delay_correlation} shows the correlation between the secondary delays for two different values of the primary delay, namely $p$=5 minutes and $p$=30 minutes. 
There is a wide spread of deviations from the solid line showing the expectation for  the case of a linear $s(p)$. This  is indicative of the multitude of strengths with which these additional, higher-$p$ effects contribute.

The challenge is now to establish in detail the relations between the degree of synchronization and the performance of the system (given by low delay propagation, i.e. robustness, and low overall transfer times, i.e. efficiency). 

On the level of our data, the main performance indicators of the system, namely the efficiency and robustness, are only indirectly accessible via the secondary delays and the buffering times. We expect that a small $s(p)$ is related to high robustness (a given perturbation $p$ induces a small effect $s$), while a small $b$ can be associated with high efficiency (during a full itinerary only a small amount of time, given by the local buffering times $b$, is accumulated upon train interchanges). 

\begin{figure}[ht]
   \begin{center}
        \includegraphics[width=0.5\textwidth]{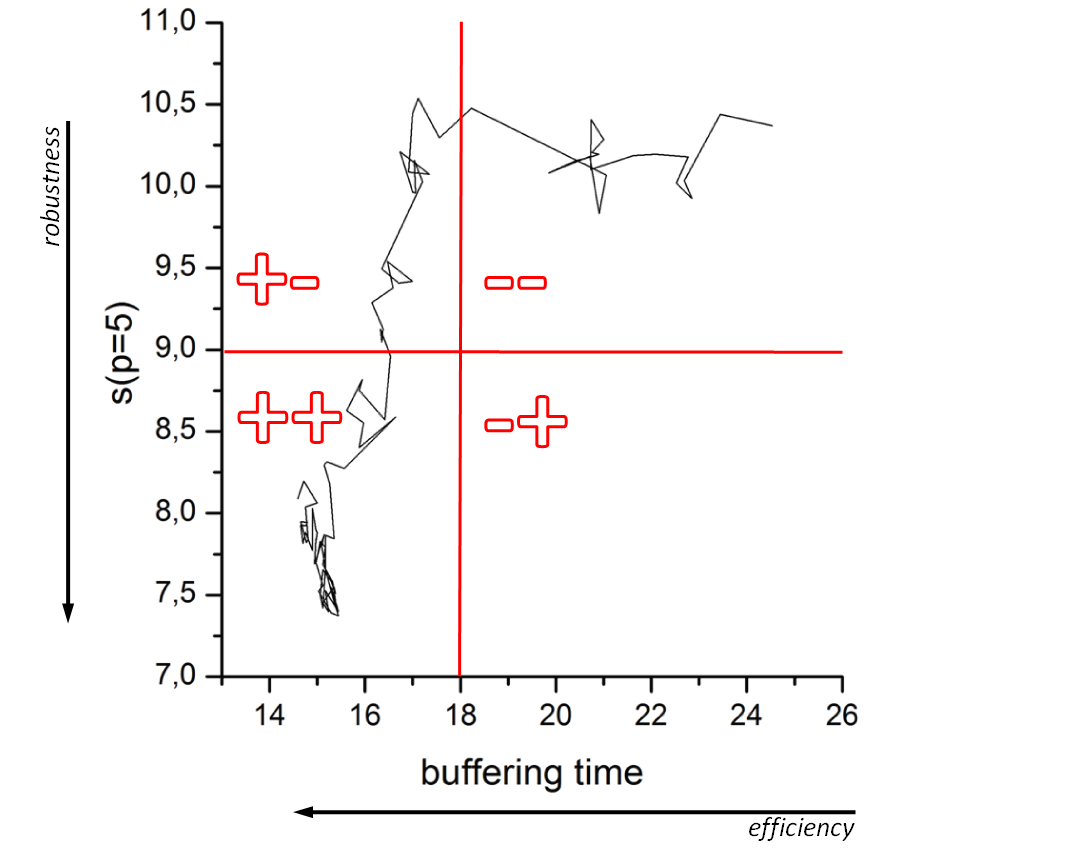}
        \caption{Secondary delay as a function of the buffering time for
          a fixed primary delay $p$ = $5$ minutes,  averaged and
          connected along the station rank, together with a phenomenological separation into quadrants according to high/low efficiency (first quadrant label) and high/low robustness (second label). \label{fig:secondary_delay_buffer}}
    \end{center}
\end{figure}

When splitting the $b$-$s(p)$--plane into four quadrants corresponding
of contributions of high/low delays and 
\linebreak buffering times -- and,
consequently, low/high 
($-$/+) performance --, one can observe very different usages (i.e., frequencies of occurrence in the data) of the quadrants:

The ++ region, which is the most efficient one as those stations are both efficient (small buffering time) and robust (small secondary delays), is most densely populated,
followed by the +$-$ region (high efficiency, low robustness). Very few stations are found in the $-$$-$ region. Interestingly we do not find stations in the $-$$+$ region. Probably those are quickly eliminated during the schedule building or avoided during the route search.

% first result: sync
In order to better understand the systematic relation between $s(p)$ and $b$ it is again helpful to use the station size as a control parameter along which local averages can be performed. The resulting curve is shown in Figure~\ref{fig:secondary_delay_buffer}. This curve displays the 
backbone systematics of the 
interplay between efficiency (inverse $b$) and robustness (inverse $s(p)$) studied from the raw data in Figure \ref{fig:secondary_delay_buffer_raw}, when using station size as an ordering parameter.

\begin{figure}[t]
   \begin{center}
        \includegraphics[width=\columnwidth]{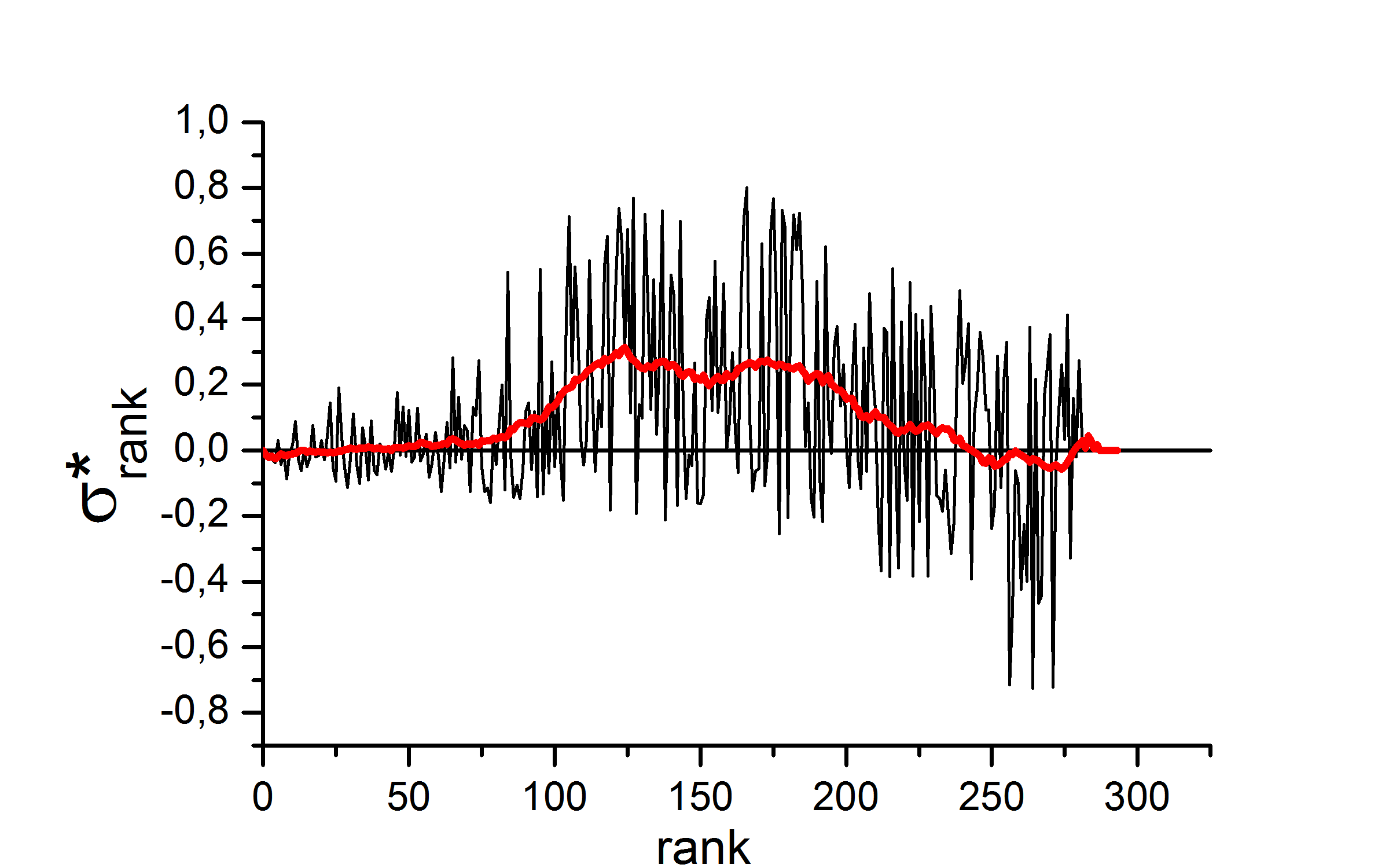}
        \caption{The synchronization indices of the stations in
          ascending order of the station
          rank in Germany with $N_R=100$ and an  averaging window of $40$\label{fig:synchronization}}
    \end{center}
\end{figure}
Figure~\ref{fig:synchronization} shows the synchronization index $\sigma_k^*$ as a function of the station rank $k$.
% smoothing 
The phase data are distorted by the mere number of A/D events. In
particular, at few A/D events large fluctuations of $\sigma$ are
induced. We therefore subtracted from each $\sigma _k$ an average
$\sigma^{(R)}_k$ over $N_R$ runs of a null model, where the same
number of A/D events has randomly been distributed in time. This
procedure yields the reduced synchronization indices
$\sigma^*_k$=$\sigma_k - \sigma^{(R)}_k$, shown as the black curve in
Figure~\ref{fig:synchronization}. 

\begin{figure}[ht]
  \centerline{\includegraphics[width=0.5\textwidth]{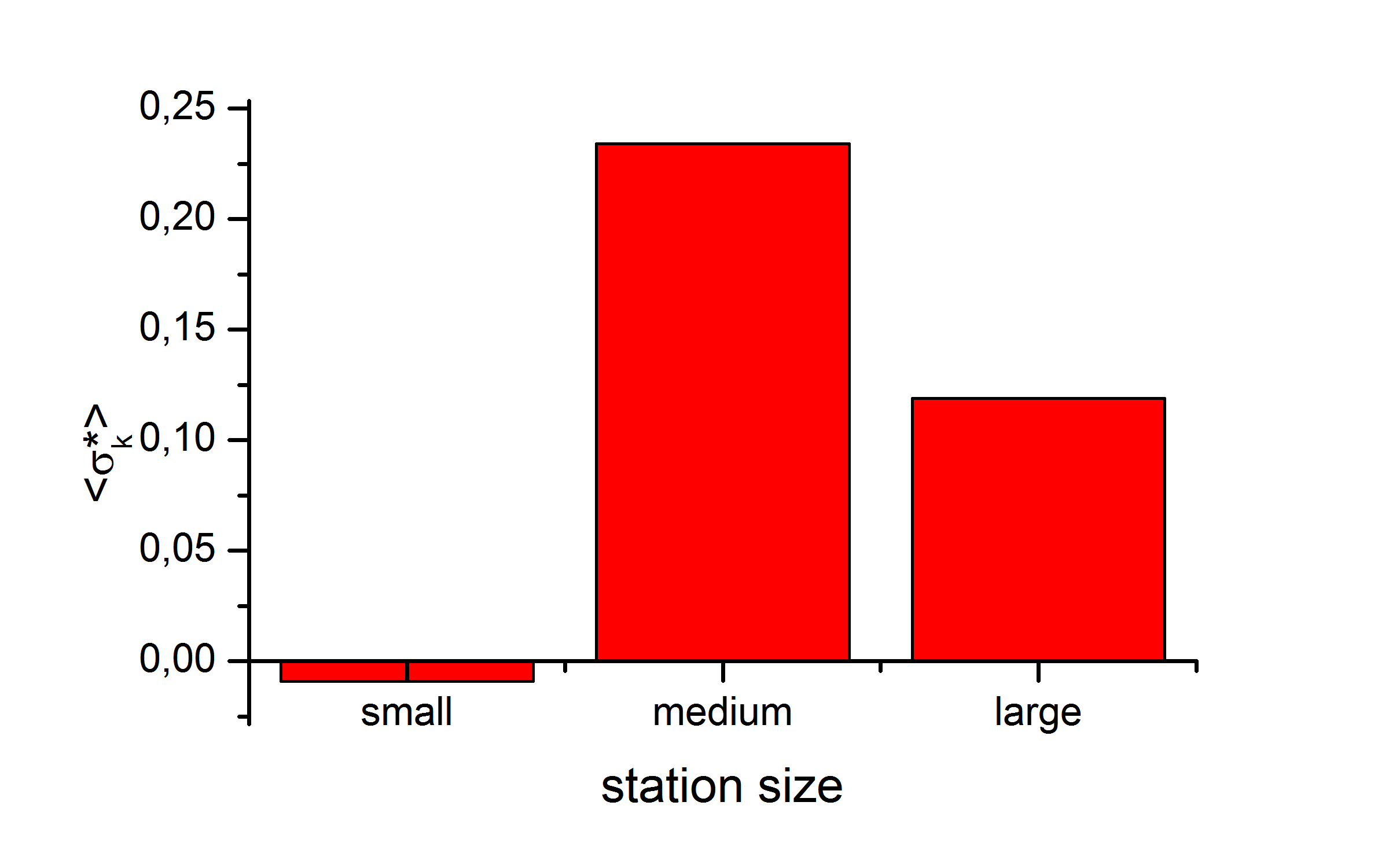}}
	\caption{Average $\sigma^*$ for small, medium and large
          stations. The rank is split at 80 and 170 A/D
          events per day, respectively. \label{fig:sync_categories}}    
   
\end{figure}

Furthermore, stations with neighboring ranks will differ (even though they are similar in size) in a variety of additional parameters. The original reduced synchronization index shows a strong local fluctuation along the rank. In order to eliminate the variation coming from these additional differences between similar-sized stations, we compute local averages over the $\sigma^*_k$. These values are shown as the red curve in Figure~\ref{fig:synchronization}. 
 Remarkably, synchronization is highest at intermediate station rank, decreasing towards both larger and 
smaller station sizes. In order to obtain this result, several processing steps of the raw data have been necessary. 
The systematic difference between the synchronization of large, small and intermediate train stations, respectively, is also seen, when average synchronization indices for each of these three categories are computed directly (Figure~\ref{fig:sync_categories}). 

\vspace*{3ex}

\begin{figure}[ht]
\centering
\subfigure[Austria]{
\includegraphics[width=0.45 \columnwidth]{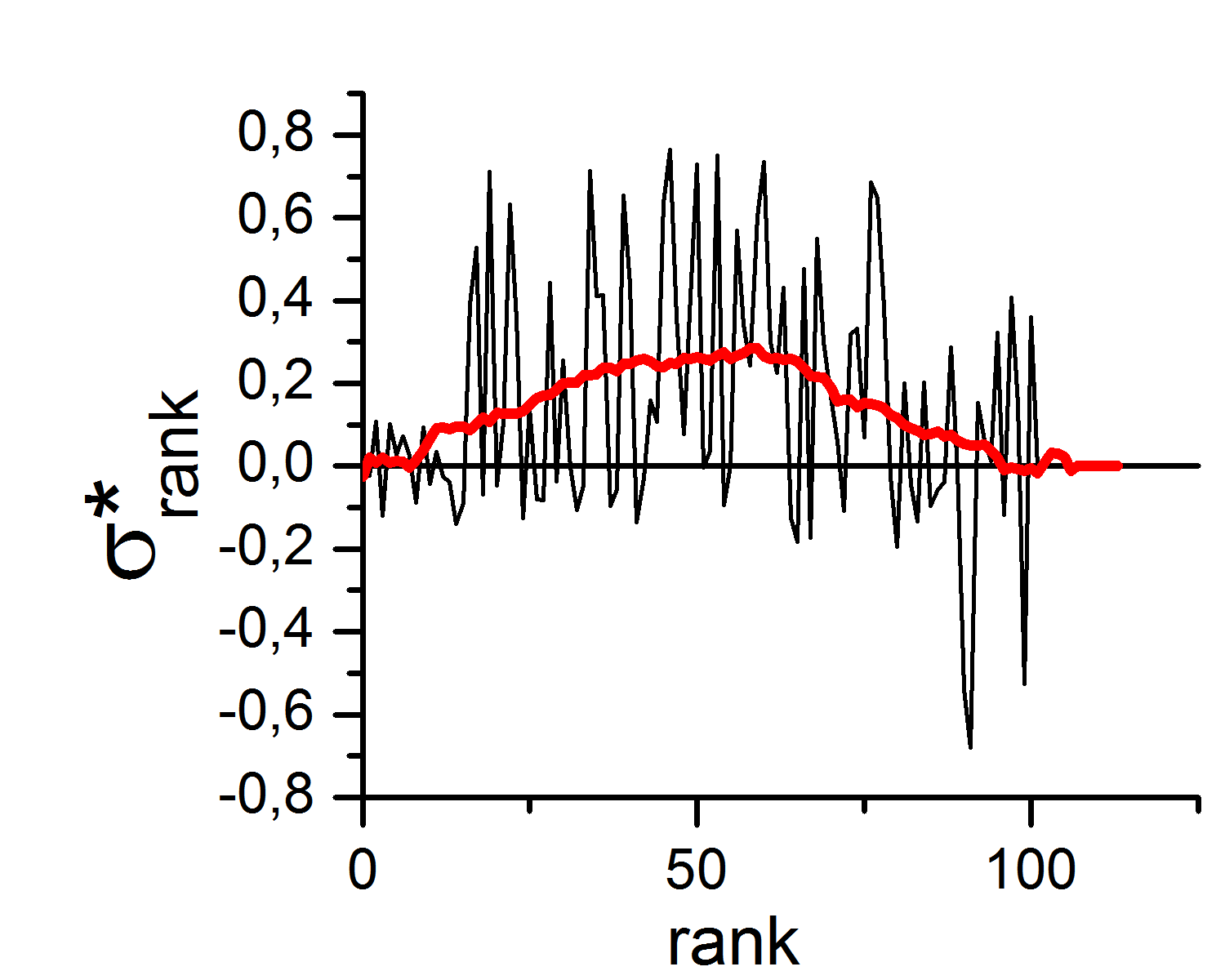}
\label{fig:austria}
}
\subfigure[France]{
\includegraphics[width=0.45 \columnwidth]{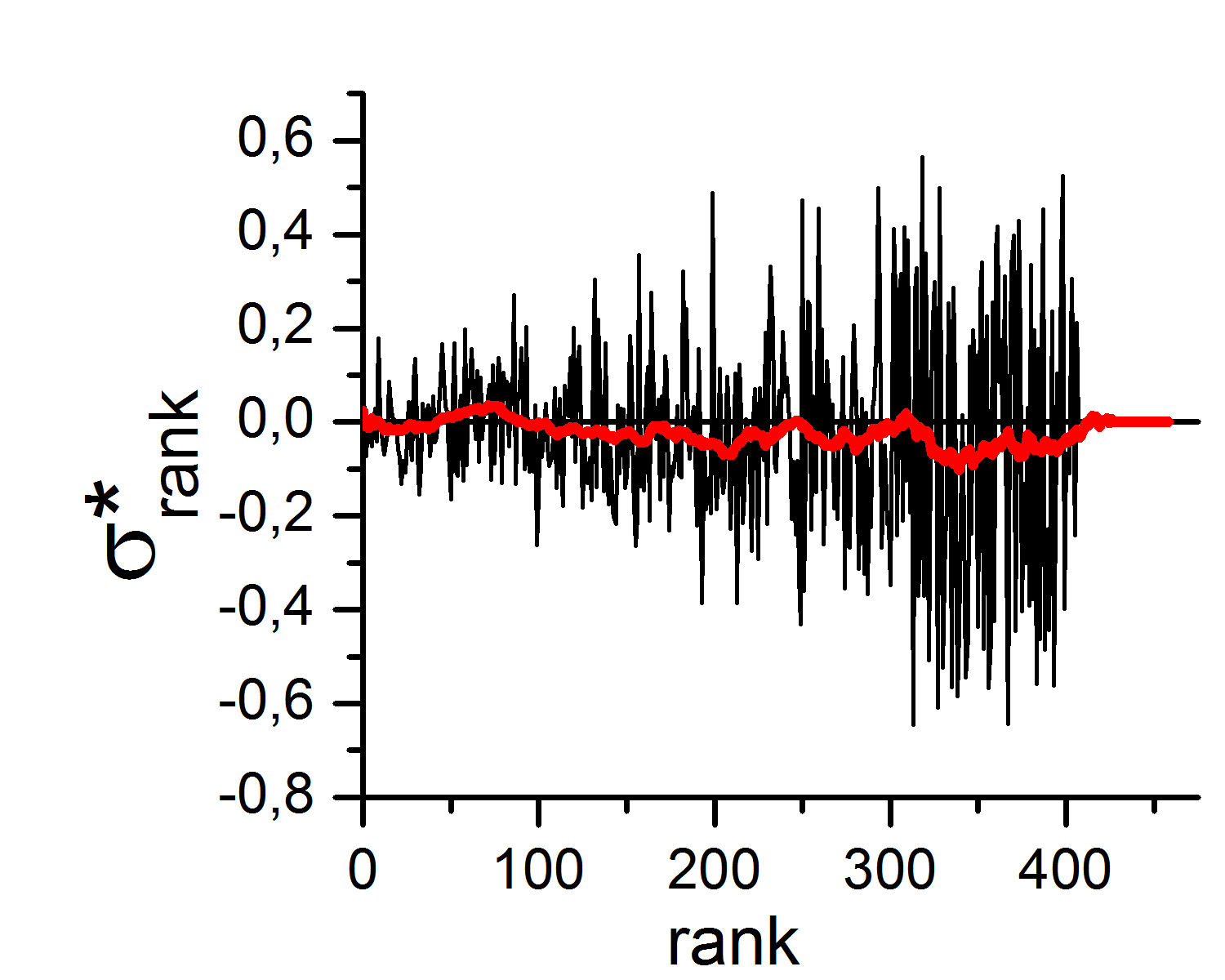}
\label{fig:france}
}
\\
\subfigure[Norway]{
\includegraphics[width=0.45 \columnwidth]{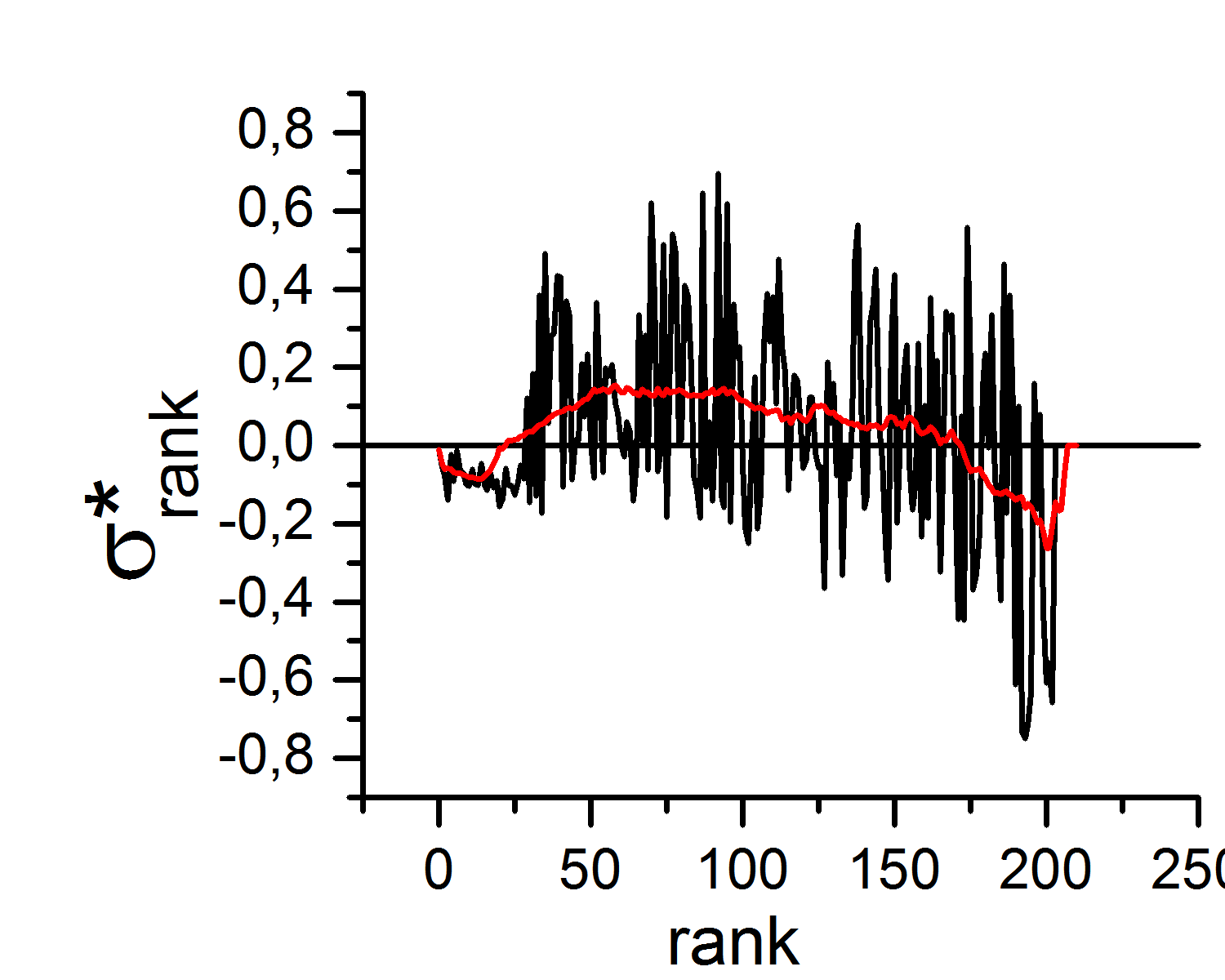}
\label{fig:norway}
}
\subfigure[Czech Republic]{
\includegraphics[width=0.45 \columnwidth]{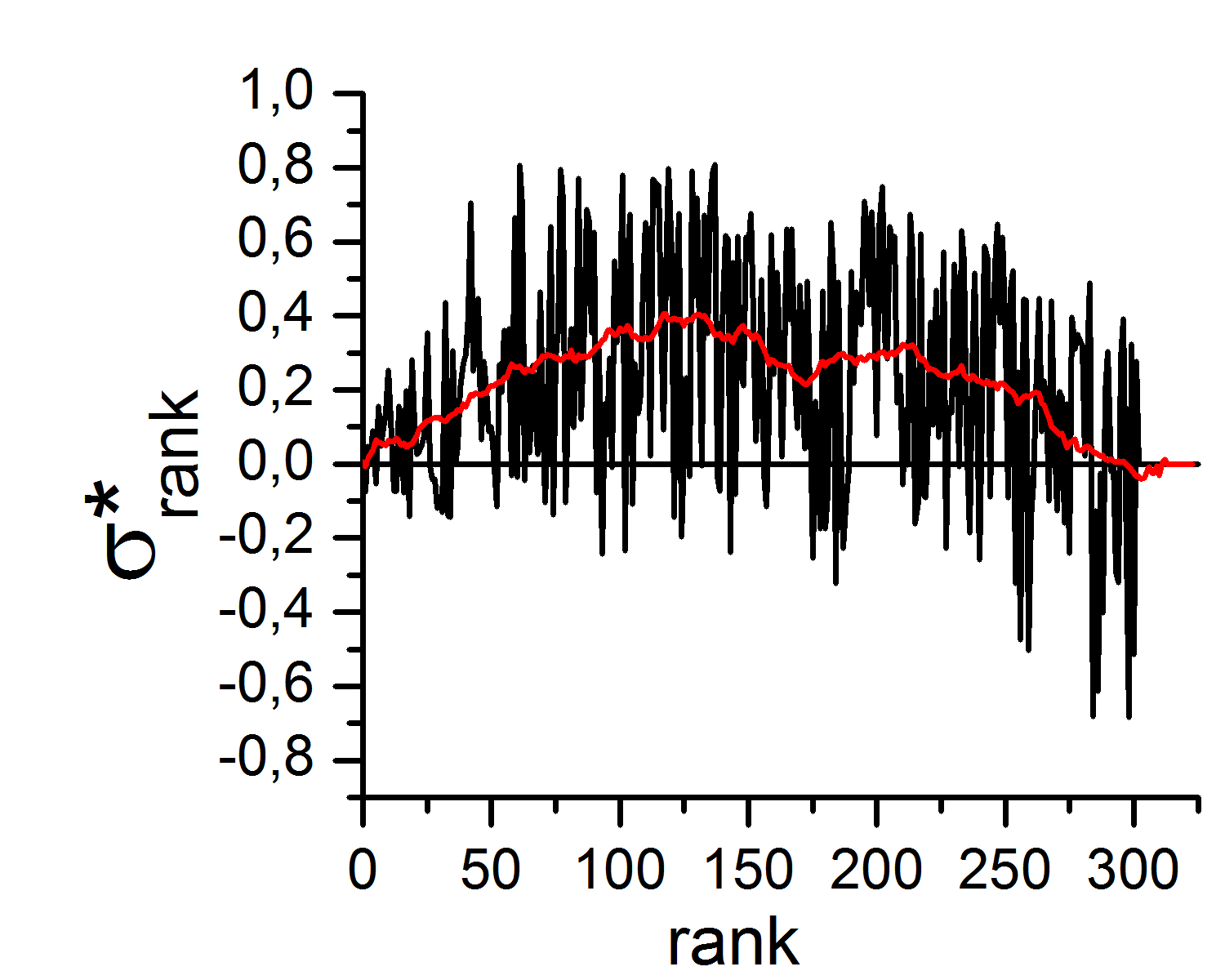}
\label{fig:czech}
}

\caption[]{The synchronization in the long-distance train connections of different European countries with $N_R=100$ and an  averaging window of $40$.\label{fig:synch_europe}%\subref{fig:subfig1} 
}
\end{figure}
%\begin{figure}[ht]
%\includegraphics[width=0.4 \columnwidth]{images/synch_austria}
%\caption{The synchronization in the Austrian long-distance train connnections. \label{fig:austria}}
%\includegraphics[width=0.4 \columnwidth]{images/synch_france}
%\caption{The synchronization in the French long-distance train connections.\label{fig:france}}
%\includegraphics[width=0.4 \columnwidth]{images/synch_norway}
%\caption{The synchronization in the Norwegian long-distance train connections.\label{fig:france}}
%\includegraphics[width=0.4 \columnwidth]{images/synch_czech}
%\caption{The synchronization in the Czech long-distance train connections.\label{fig:france}}
%\end{figure}
In order to assess, whether this elevated synchronization of  intermediate-size stations is a property of train timetables beyond this individual case, we also computed
the synchronization indices  $\sigma_k^*$ for four other counties,
Austria, France, Norway, and the Czech Republic (Figures \ref{fig:austria}-\ref{fig:czech}).
France shows only a very weak signal, where\-as the shapes of the synchronization curves in
Austria, Norway and the Czech Republic are very similar to the one observed in Germany 
(Figure~\ref{fig:synchronization}).

% second result: ++, etc. 

In the following we will show results for the inter\-dependencies of our main quantities $b_k$, $s_k(p)$ and $\sigma^*_k$. In all cases, like before, we compute local averages with respect to the rank.  
By grouping the stations according to their position in the $b$-$s(p)$--plane, Figure~\ref{fig:secondary_delay_buffer},
i.e. according to their robustness and efficiency, one can now study, whether stations from the same regions
share a common synchronization index $\sigma^*_k$.

\begin{figure}[ht]
   \begin{center}
        \includegraphics[width=0.5\textwidth]{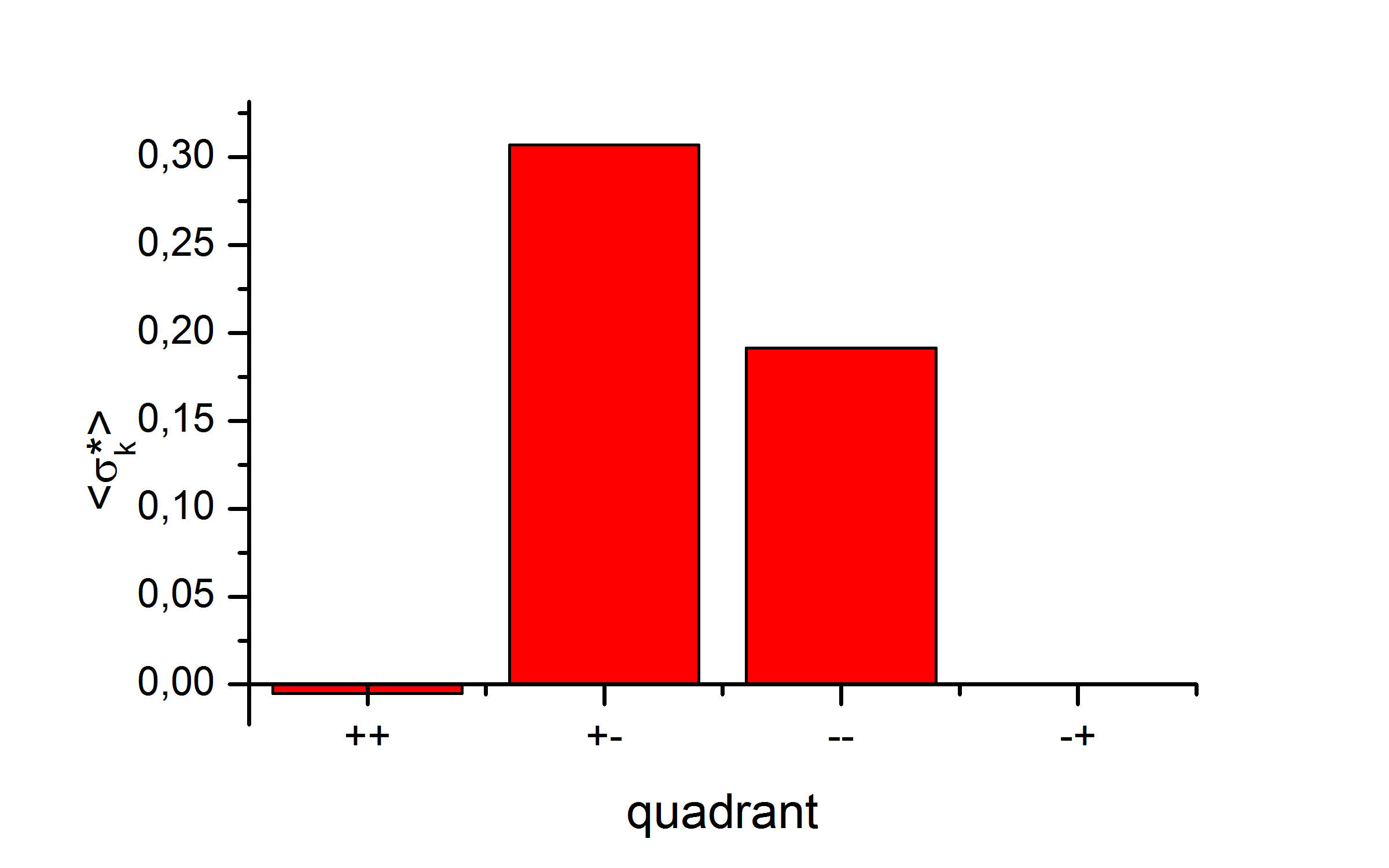}
        \caption{Average synchronization of the stations grouped
          together by robustness and efficiency according to the quadrants in Figure~\ref{fig:secondary_delay_buffer}.  \label{fig:synchronization_in_regions}}
        
    \end{center}
\end{figure}

Figure~\ref{fig:synchronization_in_regions} shows the average $\sigma$ for the
three regions containing stations. 
The stations from the most preferable region $+$$+$ show extremely
low synchronization, while those of the regions $+$$-$ and $-$$-$ are much
more synchronized.
%, see Figure~\ref{fig:synchronization_in_regions}.

\begin{figure}[ht]
   \begin{center}
        \includegraphics[width=0.5\textwidth]{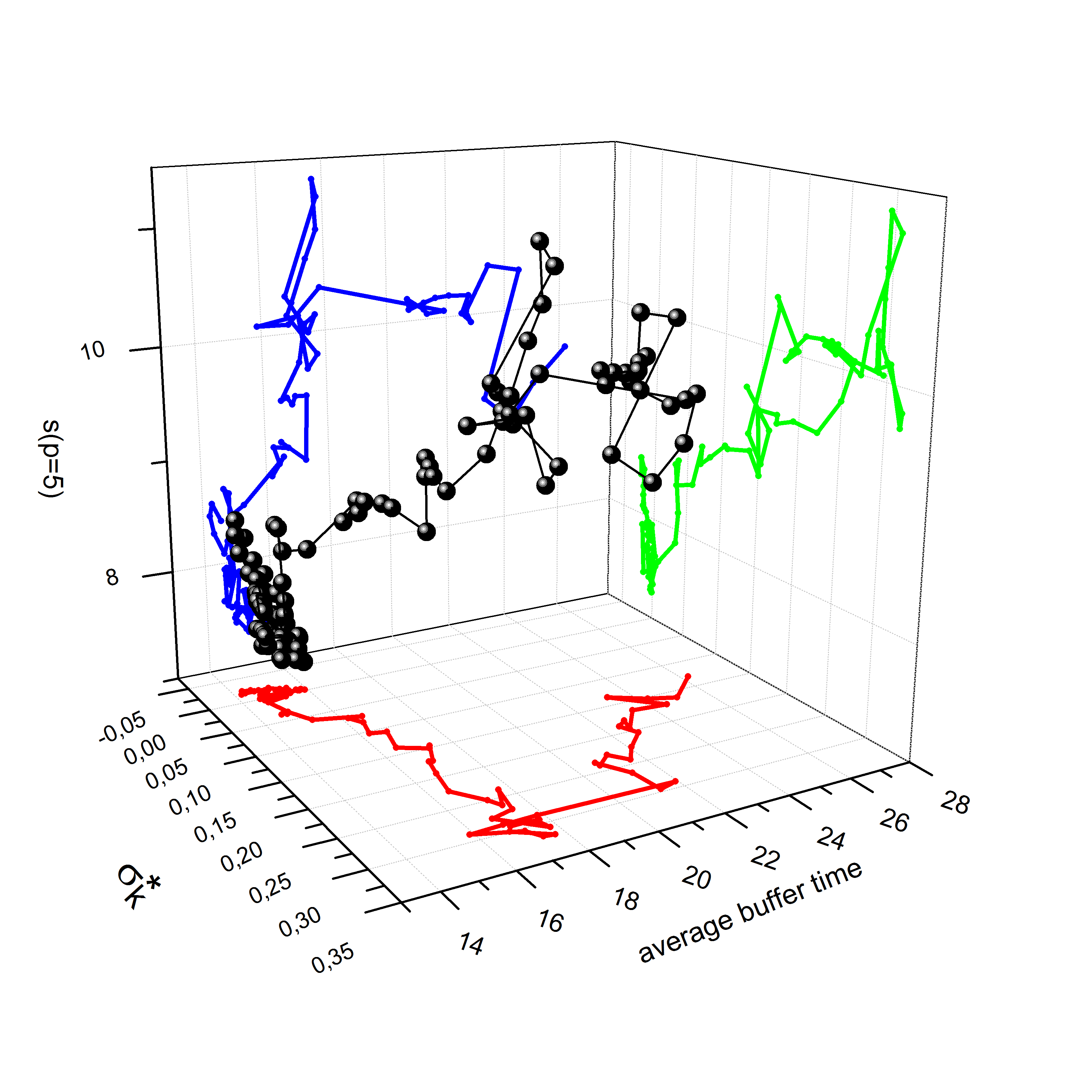}
        \caption{Dependencies of buffering time $b$, secondary delay $s(p)$, and
          synchronization index $\sigma^*_k$.
        \label{fig:synch_robust_efficient_space}}
    \end{center}
\end{figure}

 In order to show the dependencies among these quantities more
 directly, 
Figure~\ref{fig:synch_robust_efficient_space}  represents all three quantities $b_k$, $s_k(p)$ and $\sigma^*_k$, simultaneously.
The smoothing window size is set to 26. It is clearly visible that most stations are in the regions of low $\sigma^*_k$, low $b_k$ and low $s_k(p)$. Furthermore, there is a clear correlation between $s_k(p)$ and $\sigma^*_k$ and consequently, an anti-correlation between synchronization and robustness.

\section{Avalanche Model}\label{sec:avalanche}

Can the negative correlation between synchronization and robustness that we observe in the data also be understood in some minimal model of delay propagation? The general dynamical mechanism resembles some aspects of avalanches on graphs. While avalanche models are an important focus of interest in complex systems theory and in particular in the field of self-organized criticality \cite{jensen,per_bak}, we do not expect here a power-law distribution of event sizes, as the elementary processes behind delay propagation are different from the threshold-driven re-distribution schemes encoded, e.g. in the Bak-Tang-Wiesenfeld (BTW) model \cite{BTW}. 
Therefore, we adapt the general concept of an avalanche model to the dynamical needs of delay propagation. 

In our passenger delay avalanche model the dynamical variables are the accumulated delays $d_i(t)$ at node $i$ as a function of time $t$. The model has three parameters: 

(1) The transmission probability $p$ is the probability that a delay propagates from one node to an adjacent node, if the threshold is crossed. This probability describes the capacity to buffer incoming delays via transfer times. 

(2)  The amplification factor $m$ acknowledges the fact that a single train delay corresponds to multiple passenger delays; consequently, if a few incoming passengers cause a train with many outgoing passengers to wait, the total (passenger-based) delay is amplified. This parameter can be seen as the ratio of these passenger numbers, i.e. the average rate of additionally delayed passengers due to waiting. 

\begin{figure}[ht]
\includegraphics[width=\columnwidth]{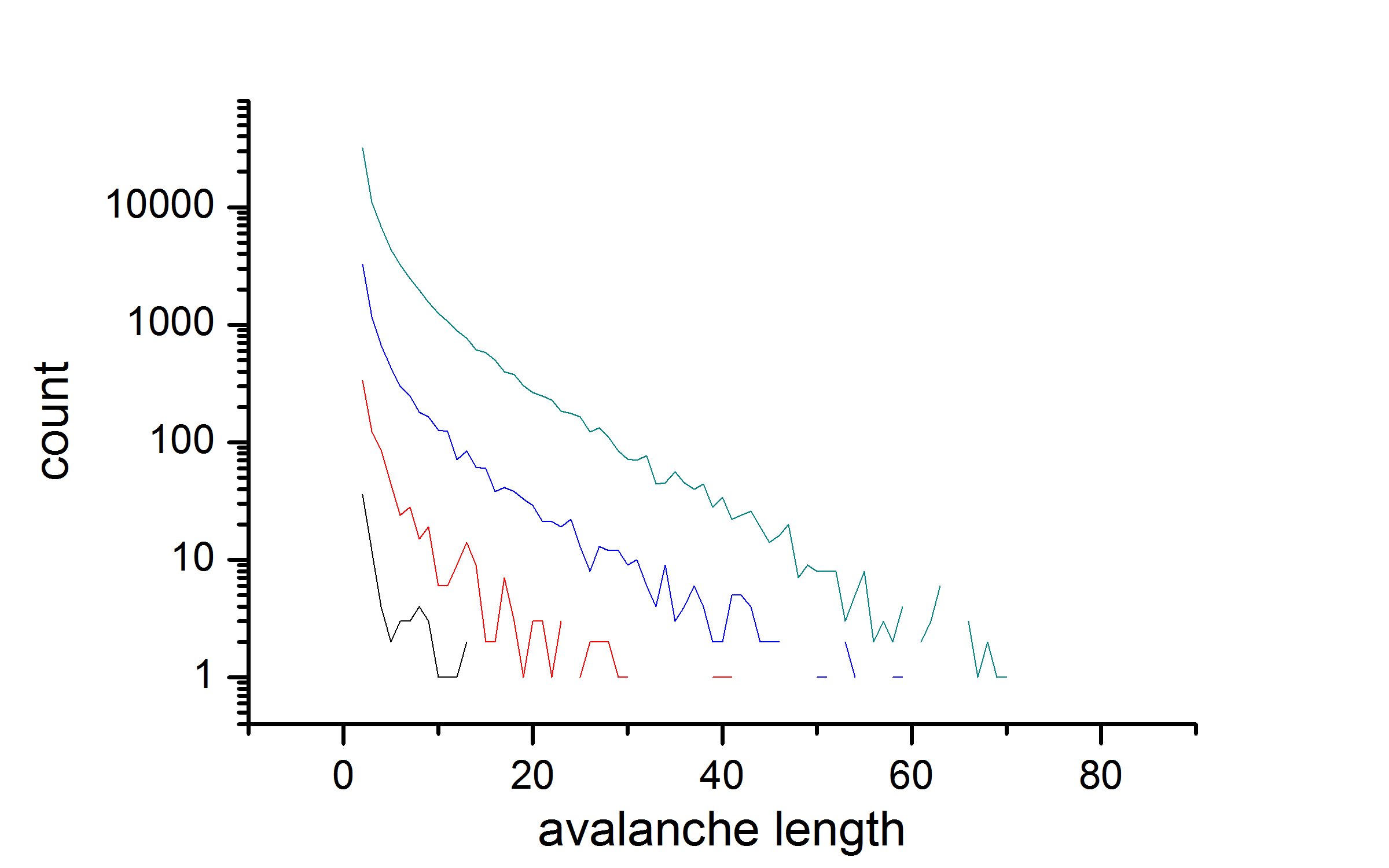}
\caption{The distribution of the avalanche lengths for stochastic and periodic drivers with a driving period of  $17$, $T=4$, $m=0.9$, $N=70$ nodes and  $M=240$ edges.   \label{fig:ava-length_dist}}
\end{figure}

(3) Delays only propagate from a node $i$ to adjacent nodes, when the delay variable $d_i(t)$ is above a threshold $T$, as we assume that only incoming delays higher than this threshold are capable of triggering delay propagation.
%In Figure~\ref{fig:PDA-model}  the general idea of this model is schematically depicted.
 Figure~\ref{fig:ava-length_dist}  shows that the tail of the size distribution of delay avalanches is exponential.

%\begin{figure}[t]
%\vspace{-30pt}
%\includegraphics[width=.85 \textwidth]{images/model}
%\vspace{-280pt}
%\caption{The passenger-delay-avalanche model (PDA-model). \label{fig:PDA-model}}
%\end{figure}

In order to analyze the relation between synchronization and robustness within this model, we compare the average avalanche length for a system, where a single node is periodically driven, with the case of a stochastically driven node.

\begin{figure}[ht]
\includegraphics[width=\columnwidth]{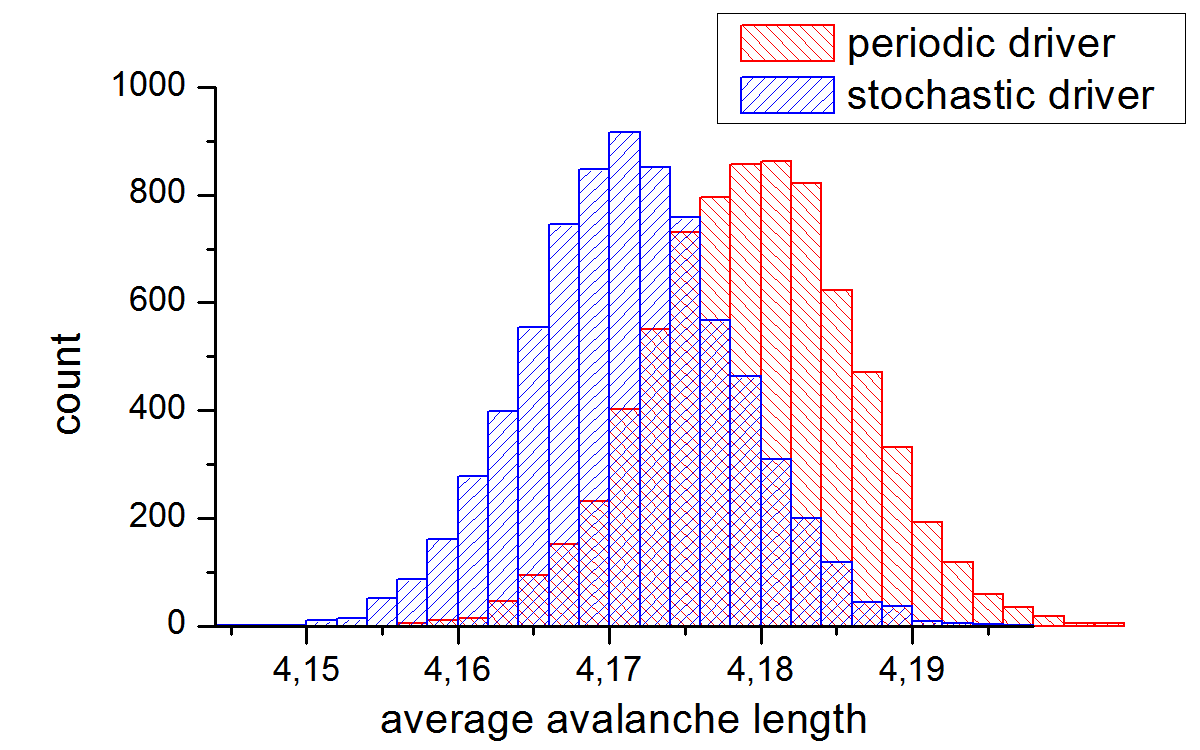}
\caption{The average avalanche length for stochastic and periodic drivers with a driving period of  $17$, $T=4$, $m=0.9$, $N=70$ nodes and  $M=240$ edges.   \label{fig:ava-length}}
\end{figure}

We assume that the gradual insertion of delay units into the system corresponds to incoming delays from other parts of the network entering the sub-network, which is here studied in detail. A periodic insertion of such delay units then corresponds to highly synchronized arrival/de\-parture (A/D) events (as only those can give rise to periodic delays), while the stochastically driven node represents the typical pattern of incoming delays for a station with less synchronized A/D events. 

Figure~\ref{fig:ava-length} shows the distribution of these average ava\-lanche lengths for the two cases. The periodically driven node (high synchronization of A/D events) coincides with a high average size of the delay avalanches (i.e. higher vulnerability or lower robustness), while the stochastically driven node (low synchronization of A/D events) displays a lower average size of delay avalanches (and therefore a higher robustness). This is in agreement with the relationship discovered in the real train connection data studied in the previous section.

\section{Conclusions}\label{sec:summary}

In this work we have studied railway timetables from a novel and yet unexplored view, namely that of phase synchronization. For our analysis we investigated the German long-distance train timetable with respect to three distinct properties: robustness, efficiency and phase synchronization. 

The robustness reflects the stability of the system to small
perturbations, while efficiency is related to short accumulated waiting times per
train route. 
These two properties have been evolved over the years by gathered
experience and heuristic optimization. 

When we consider the arrival and departure events of all trains at a given station over a period of time, 24 hours for example, we can translate those events into phases. Summing over all different phases we can compute a synchronization index for each station. Then, by exhaustive simulation we produce a primary delay at each station and record the induced secondary delays. Our results show a clear and surprising correlation between the synchronization index of a station, its robustness and efficiency. 

In the Introduction, we have discussed the difference between car
traffic in cities and the impact of traffic light synchronization on
the one side and railway timetables on the other.
It would
 be interesting to compare these two types of traffic in detail, to
 quantitatively analyze the number of directions (node degrees) 
 in the context of an
 effective dimension, and in particular to study the complexity
 (given, e.g., by the pattern of elementary decisions needed to
 specify the path) of a typical path in the train network compared to
 the car traffic case. A suitable methodology could be the framework
 developed in \cite{PhysRevE.72.046117}.

The balance between this antagonistic pair of requirements, efficiency
and robustness, is of broad interest across many disciplines, ranging
from industrial production to biological processes. Lack of robustness
due to too high efficiency is sometimes called the systemic risk, which
has recently been discussed from a theoretical perspective, for
example for complex economical systems 
(see \cite{Battiston-et-al08,Battiston-et-al2007,LorenzBattistonSchweitzer09}).
%Battiston et al. 2007; 2008; Lorenz et al. 2009).

Starting from an information-theoretical description of resilience in
ecology, Ulanowicz et al.~\cite{Ulanowicz-etal09}
%(2008)
 could establish quantitative links between 
 sustainability, efficiency and investments in diversity. This general
 framework has been employed to analyze the current bank crisis from a
 ecosystem perspective \cite{LietaerUlanowiczGoerner09}.
 %(Lietaer et al. 2009).
We believe that a quantitative view on synchronization of
arrival/departure events in the network of long-distance train
connections, as presented here, can similarly serve as a starting
point for a theoretical understanding, and subsequently systemic 
optimization, of the balance between efficiency and robustness for 
such timetables underlying public transportation.

For biological processes this balance between efficiency and robustness has
 been explored in a multitude of ways resorting to both analysis of 
experimental data and the  mathematical modeling of cellular
processes. 
Motivated by graph theory and nonlinear dynamics, 
an influential trend in systems biology at the moment is to relate 
robustness to small regulatory devices 
\cite{Alon07,BrandmanMeyer08},
%(Alon 2007, Brandman and Meyer 2008), 
serving e.g. as a noise buffer 
or providing a suitable amount of redundancy for maintaining systemic 
function even under perturbations. In particular such relations
between the architecture of regulatory devices and dynamical functions
have been worked out for circuits of negative feedback loops 
\cite{PigolettiKrishnaJensen07},
%(Pigolotti et al. 2007), 
for feedforward loops as noise filtering
devices in gene regulation 
\cite{Alon07,Shen-Orr-et-al02},
%(Shen-Orr 2002; Alon 2007), 
for interlinked feedback loops acting on different time scales 
\cite{Brandman-et-al05},
%(Brandman et al. 2005), 
for a particular composition of regulatory 
units 
\cite{Milo-et-al04}
%(Milo et al. 2004) 
and their relation to robustness
\cite{KlemmBornholdt05,Kaluza-et-al07,KaluzaMikhailov07,Kaluza-et-al08},
%(Klemm and Bornholdt 2005, Kaluza et al. 2007, 2008; 
%Kaluza and Mikhailov 2007), 
for number of positive and negative 
feedback loops in regulatory circuits
\cite{KwonCho08}.
% (Kwon and Cho 2008).

It could well be that in the network of long-distance train
connections such small, motif-like network components serve as
mediators between synchronization, reliability and efficiency. 
Exploring the involvement of network topology in shaping this
relationship is one of our principal goals 
in the continuation of the work presented here.

% \begin{figure}[ht]
%    \begin{center}
%         \includegraphics[width=0.45\textwidth]{images/rank+avg_buf+S(rank-31)}
%         \caption{The space \label{fig:exchanges}}
%      \end{center}
% \end{figure}

%
% BibTeX users please use
\bibliographystyle{plain}
\bibliography{trainsync}
% \bibliographystyle{}
% \bibliography{}
%\begin{thebibliography}{}
%Author, \textit{Book title} (Publisher, place year) page numbers
% etc
%\end{thebibliography}

\end{document}